\documentclass[11pt,twoside,a4paper,cmspaper,final,collab]{cms-tdr}
\def\svnVersion{49661c4-D}\def\svnDate{2026/10/28}\def\cmsCernNoTag{CERN-EP-2026-279}\def\cmsCernDate{\today}\def\cmsMessage{Submitted to the Journal of High Energy Physics}
\begin{document}\cmsNoteHeader{EXO-24-015}

\newcommand{\Pellprpr}{\HepParticle{\Pell}{}{\prime\prime}\xspace}
\newcommand{\PGnellpr}{\HepParticle{\PGn}{\Pellpr}{}\xspace}
\newcommand{\keVns}{\ensuremath{\text{ke\hspace{-.08em}V}}\xspace}

\newcommand{\pp}{\ensuremath{\Pp\Pp}\xspace}

\newcommand{\mN}{\ensuremath{m_{\PN}}\xspace}
\newcommand{\tauN}{\ensuremath{\tau_{\PN}}\xspace}
\newcommand{\VN}{\ensuremath{\abs{V}}\xspace}
\newcommand{\VNsq}{\ensuremath{\VN^2}\xspace}
\newcommand{\VNe}{\ensuremath{\abs{V_{\PN\Pe}}}\xspace}
\newcommand{\VNm}{\ensuremath{\abs{V_{\PN\PGm}}}\xspace}
\newcommand{\VNt}{\ensuremath{\abs{V_{\PN\PGt}}}\xspace}
\newcommand{\VNmsq}{\ensuremath{\VNm^2}\xspace}

\newcommand{\muzero}{\ensuremath{\PGm_0}\xspace}
\newcommand{\muone}{\ensuremath{\PGm_1}\xspace}
\newcommand{\mutwo}{\ensuremath{\PGm_2}\xspace}
\newcommand{\muonetwo}{\ensuremath{\PGm_{1/2}}\xspace}
\newcommand{\muprime}{\ensuremath{\PGm^\prime}\xspace}

\newcommand{\abseta}{\ensuremath{\abs{\eta}}\xspace}
\newcommand{\ptmumu}{\ensuremath{\pt^{\text{pair}}}\xspace}
\newcommand{\DRmumu}{\ensuremath{\DR(\muone,\mutwo)}\xspace}
\newcommand{\mmumumu}{\ensuremath{m(\muzero,\muone,\mutwo)}\xspace}
\newcommand{\Dphimumu}{\ensuremath{\Dphi(\muzero,\muonetwo)}\xspace}
\newcommand{\mZveto}{\ensuremath{m(\muzero,\muprime)}\xspace}
\newcommand{\Lxy}{\ensuremath{L_{xy}}\xspace}
\newcommand{\Lxysig}{\ensuremath{\Lxy/\sigma_{\Lxy}}\xspace}
\newcommand{\Irel}{\ensuremath{I_{\text{rel}}}\xspace}
\newcommand{\Irelch}{\ensuremath{I_{\text{rel}}^{\text{ch}}}\xspace}
\newcommand{\Ich}{\ensuremath{I_{\text{ch}}}\xspace}
\newcommand{\Deta}{\ensuremath{\Delta\eta}\xspace}
\newcommand{\Dphi}{\ensuremath{\Delta\phi}\xspace}
\newcommand{\pSV}{\ensuremath{p_{\text{SV}}}\xspace}
\newcommand{\chisq}{\ensuremath{\chi^2}\xspace}
\newcommand{\costheta}{\ensuremath{\cos\theta(\muone,\mutwo)}\xspace}

\newcommand{\Bpr}{\ensuremath{\text{B}^\prime}\xspace}
\newcommand{\Bprpr}{\ensuremath{\text{B}^{\prime\prime}}\xspace}
\newcommand{\Dpr}{\ensuremath{\text{D}^\prime}\xspace}
\newcommand{\Dprpr}{\ensuremath{\text{D}^{\prime\prime}}\xspace}

\newcommand{\numA}{\ensuremath{n_{\text{A}}}\xspace}
\newcommand{\numB}{\ensuremath{n_{\text{B}}}\xspace}
\newcommand{\numC}{\ensuremath{n_{\text{C}}}\xspace}
\newcommand{\numD}{\ensuremath{n_{\text{D}}}\xspace}
\newcommand{\numBpr}{\ensuremath{n_{\text{\Bpr}}}\xspace}
\newcommand{\numBprpr}{\ensuremath{n_{\text{\Bprpr}}}\xspace}
\newcommand{\numDpr}{\ensuremath{n_{\text{\Dpr}}}\xspace}
\newcommand{\numDprpr}{\ensuremath{n_{\text{\Dprpr}}}\xspace}

\newcommand{\TF}{\ensuremath{\text{TF}}\xspace}
\newcommand{\CP}{\ensuremath{C\!P}\xspace}

\cmsNoteHeader{EXO-24-015}
\title{Search for long-lived heavy neutral leptons in signatures with one prompt muon and a secondary vertex formed by two non-prompt muons in proton-proton collisions at \texorpdfstring{$\sqrt{s}=13$ and 13.6\TeV}{sqrt(s)=13 and 13.6 TeV}}

\date{\today}

\abstract{A search for heavy neutral leptons (HNLs) in the decay of \PW bosons is performed in final states with three muons, with one originating promptly from the primary interaction vertex and the other two stemming from a common secondary vertex. The analyzed proton-proton collision data correspond to an integrated luminosity of 138\fbinv recorded at a centre-of-mass energy of 13\TeV in 2016--2018, and of 34.6\fbinv recorded at 13.6\TeV in 2022. The analysis strategy is optimized for large HNL decay lengths by requiring highly non-prompt muons forming a secondary vertex far from the beam line, and hence reconstructed only in the muon system of the CMS experiment at the CERN LHC. This enhances the analysis reach in the high HNL lifetime region, corresponding to low HNL masses (2--4\GeV) and small mixing parameters with the standard model neutrinos ($10^{-6}$--$10^{-5}$), where prompt and tracker-based non-prompt analyses have limited acceptance. No excess over the background prediction is found. Under the Majorana HNL hypothesis, exclusion limits at 95\% confidence level are set on the HNL mixing parameter as functions of the HNL mass in the range of 1--5\GeV.}

\hypersetup{%
pdfauthor={CMS Collaboration},%
pdftitle={Search for long-lived heavy neutral leptons in signatures with one prompt muon and a secondary vertex formed by two non-prompt muons at in proton-proton collisions sqrt(s)=13 and 13.6 TeV},%
pdfsubject={CMS},%
pdfkeywords={CMS, HNLs, Long-lived}}

\maketitle

\section{Introduction}

The discovery of neutrino oscillations~\cite{Nu_oscillations_T2K, Nu_oscillations_SNO, Nu_oscillations_Opera} establishes that neutrinos have non-zero masses, providing possible evidence for physics beyond the standard model (SM).
Constraints from cosmological observations~\cite{chacko_cosmological_2020, m_nu_upperLimits_cosmology} and direct measurements~\cite{formaggio_direct_2021, m_nu_upperLimits_Katrin} indicate that neutrino masses are several orders of magnitude smaller than those of other fermions.
Several theoretical extensions of the SM have been proposed that explain both the non-zero mass values and the anomalous mass scale of the neutrino sector through the introduction of massive right-handed counterparts of the SM neutrinos, referred to as heavy neutral leptons (HNLs).
By including right-handed neutrinos in the theory, it is possible to write a gauge-invariant neutrino mass term, where the small mass values of the lightest neutrino physical states naturally originate from the seesaw mechanism~\cite{minkowski_e_1977, Yanagida:1979as, Seesaw_typeI_Schechter, Mohapatra:1979ia}.
In addition, the existence of HNLs could explain other unresolved questions in particle physics and cosmology.
For example, a long-lived HNL with a mass in the \keVns range would be a viable dark matter candidate~\cite{Dodelson:1993je, Dolgov:2000ew, nuMSM_paper1, nuMSM_paper2}, and the potential presence of \CP-violating effects in the HNL sector could contribute to the matter-antimatter asymmetry in the universe~\cite{nuMSM_paper1, boyarsky_role_2009, Baryogenesis_nuOscillations}.
Right-handed neutrinos are singlets under the SM gauge group, and thus do not participate in the electroweak or strong interactions.
However, HNLs can still be produced in charged- and neutral-current weak interactions via mixing with the SM neutrinos.

\begin{figure}[!t]
\centering
\includegraphics[width=0.49\textwidth]{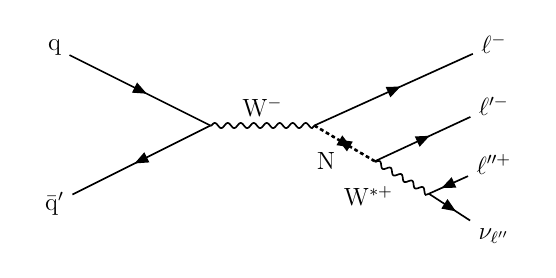}%
\hfill%
\includegraphics[width=0.49\textwidth]{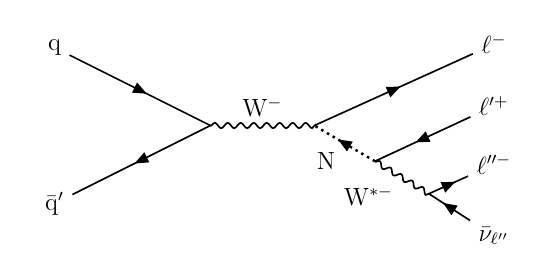} \\
\includegraphics[width=0.49\textwidth]{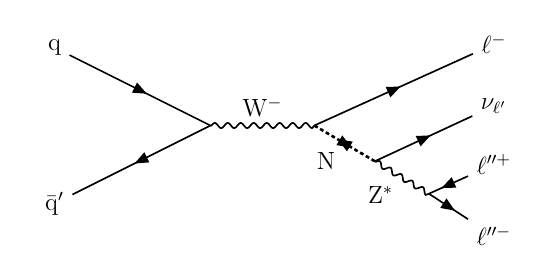}%
\hfill%
\includegraphics[width=0.49\textwidth]{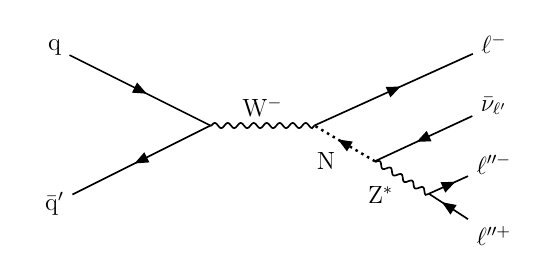}
\caption{%
    Feynman diagrams for the production of an HNL (denoted as \PN) through its mixing with an SM neutrino, resulting in a final state with three charged leptons and a neutrino.
    The symbols \PQq and $\PAQq^{\prime}$ denote the incoming quark and antiquark, respectively.
    In the notation used in this figure, \Pell originates from the primary \PWm boson decay, \Pellpr from the HNL decay directly, and \Pellprpr from the leptonic decay of the virtual electroweak boson.
    The HNL decay is mediated by a virtual \PW boson (\PWst) in the upper row and by a virtual \PZ boson (\PZst) in the lower row.
    On the left, the HNL is assumed to be a Majorana neutrino, allowing \Pell and \Pellpr to have the same charge in the diagram mediated by the \PWst boson, with consequent lepton number violation.
    On the right, instead, the HNL can be either a Dirac or a Majorana particle.
}
\label{fig:feynman_diagrams}
\end{figure}

This search targets HNL production in decays of \PW bosons produced in proton-proton (\pp) collisions at the CERN LHC.
The studied channel involves the decay of a \PW boson into a charged lepton and an HNL, with a subsequent decay of the latter into two charged leptons and a neutrino, mediated either through a \PW or a \PZ boson.
Hence, the resulting final state consists of three charged leptons and a neutrino.
Example Feynman diagrams for this fully leptonic final state are shown in Fig.~\ref{fig:feynman_diagrams}.
For a Majorana HNL, the decay mediated by a \PW boson can violate lepton number, allowing the charged lepton \Pell from the \PW boson decay and the lepton \Pellpr from the HNL decay to have the same charge. This configuration is forbidden for a Dirac HNL, for which lepton number is conserved. The three-lepton final state considered here can arise in both cases.
The flavour of the final-state leptons in the diagrams in Fig.~\ref{fig:feynman_diagrams} depends on the mixing between the HNL and SM neutrinos, which is described by the three mixing parameters \VNe, \VNm, and \VNt, where the symbol \PN represents the HNL.
The HNL can either couple to a single neutrino family (\ie only one of the mixing parameters is different from zero) or to multiple families (\ie at least two of \VNe, \VNm, and \VNt are simultaneously non-zero with variable relative strength).
In the first case, the charged lepton \Pell from the \PW boson decay and the lepton \Pellpr (or \PGnellpr) from the HNL decay always belong to the same family, and lepton flavour is conserved.
In the case of mixing with multiple generations, instead, lepton flavour can be violated, \ie $\Pell\neq\Pellpr$.

{\tolerance=800
This search focuses on the final state with three muons, where the two muons from the HNL decay are highly non-prompt (and referred to as ``displaced'' in what follows).
In contrast to electrons, which shower in the electromagnetic calorimeter (ECAL) and require tracker information for their momentum reconstruction, displaced muons can be reconstructed using only the outer muon system, providing a signature of HNLs decaying in the outer layers of the inner tracker or beyond.
The three-muon channel provides maximal sensitivity to the lepton-flavour-conserving scenario of single-family mixing with muon neutrinos: \mbox{$\VNe:\VNm:\VNt=0:1:0$}.
In scenarios with simultaneous mixing with other lepton families, the HNL production and decay rates are distributed among several final-state flavour combinations, reducing the probability of the three-muon final state and thus weakening the sensitivity of this search relative to the above benchmark.
\par}

Both the production cross section and the HNL lifetime depend on its mass \mN and the coupling strength $\VNsq=\VNe^2+\VNm^2+\VNt^2$.
The proper lifetime \tauN is related to these parameters by $\tauN\propto\mN^{-5}\VN^{-2}$, resulting in a rapid increase of \tauN in the low-mass and low-coupling region.
Hence, the signature of HNL decays at low \mN (of the order of 10\GeV and lower) is significantly affected by their long lifetime, which needs to be accounted for in the simulation and in the interpretation of the results.
The long lifetime is also a powerful handle to suppress backgrounds from SM processes, which typically have prompt decays.
In particular, the decay products of light HNLs with long lifetimes will originate from a secondary vertex (SV) that is spatially displaced from the hard interaction, or primary vertex (PV), and can be distinguished from it.
Tracker-based searches require the displaced charged-particle tracks to be reconstructed in the silicon tracker and consequently have limited acceptance for HNL transverse decay lengths above 1\unit{m}. The present search instead targets transverse decay lengths of approximately 50--400\unit{cm}, covering decays in the outer layers of the inner tracker or beyond, by selecting two muons that can only be reconstructed using information from the CMS muon system. Such highly displaced dilepton vertices provide a clean signature of long-lived HNL decays and probe the long-lifetime region, corresponding in this search primarily to $\mN\lesssim5\GeV$ and small \VNsq ($10^{-6}$--$10^{-5}$).

Extensive searches have been conducted, based on the experimental signatures of HNL models~\cite{delAguila:2008cj, HeavyN_model_theory, Tello:2010am, Deppisch:2015qwa, Cai:2017mow, Pascoli:2018heg, Alimena:2019zri, Bondarenko:2019tss, Abdullahi:2022jlv, Antel:2023hkf}.
For $\mN\lesssim2\GeV$, the most stringent HNL constraints come from fixed-target experiments at extracted beam lines, such as CHARM~\cite{CHARM:1985nku}, NA62~\cite{NA62:2021bji}, T2K~\cite{T2K:2019jwa}, and E949~\cite{E949:2014gsn}, where long-lived HNLs are produced in the decays of pions, kaons, and heavy-quark hadrons.
Several searches for HNLs have also been performed at the LHC by the ATLAS, CMS, and LHCb Collaborations~\cite{CMS:EXO-11-076, CMS:EXO-12-057, ATLAS:2015gtp, CMS:EXO-14-014, CMS:EXO-17-012, CMS:EXO-17-028, ATLAS:2019kpx, LHCb:2020wxx, CMS-EXO-20-009, ATLAS_HNL_3l_Run2, CMS:EXO-21-003, ATLAS:2023tkz, CMS:2023jqi, CMS:2024ake, CMS:EXO-22-011, CMS:2024ita, ATLAS:2024rzi, CMS:2024bni, CMS:2024hik, ATLAS:2024fcs, ATLAS:2025uah, ATLAS:2025lva, LHCb:2025ymr}.
These searches target various production mechanisms and final states, and some of them probed the same \mN range considered here.
The CMS Collaboration reported on a search for HNLs in final states with three charged leptons (electrons or muons) and displaced vertices~\cite{CMS-EXO-20-009}, using \pp collision data recorded in 2016--2018 at $\sqrt{s}=13\TeV$.
Similar searches were reported by the ATLAS Collaboration, using data recorded in 2015--2018~\cite{ATLAS_HNL_3l_Run2, ATLAS:2025uah}.
In both cases, the presence of an SV leads to the extension of the sensitivity of previous searches to the low-mass, low-coupling regime.

Other CMS analyses using the 13\TeV LHC data searched for HNLs produced via \PW bosons and decaying semileptonically.
The considered final states include those with one prompt lepton associated with at least one displaced hadronic jet~\cite{CMS:2023jqi}, an SV formed by a lepton and a jet~\cite{CMS:2024hik}, and large particle showers in the muon detectors due to highly displaced HNL decays~\cite{CMS:2024ake}.
Additionally, a search for HNLs produced in semileptonic \PB meson decays was performed using a special data sample enriched in \PB mesons~\cite{CMS:2024ita}.
A comprehensive overview of the status of HNL searches in CMS is provided in Ref.~\cite{CMS:2024bni}.

The final state targeted by this search consists of one prompt muon and two highly displaced muons from an HNL decay reconstructed using only information from the CMS muon detectors.
The two displaced muon candidates are required to form an SV and to not match any track in the tracker system.
This latter requirement has two consequences: it extends the analysis acceptance to larger HNL decay lengths, and it enables the suppression of most of the SM background processes that exhibit prompt or moderately displaced signatures.
In addition, by considering displaced muons that are not associated with tracker information, the present search is complementary to the one reported in Ref.~\cite{CMS-EXO-20-009}.

The search utilizes two data sets recorded at different centre-of-mass energies.
The data set corresponding to an integrated luminosity of 138\fbinv was recorded in 2016--2018 at $\sqrt{s}=13\TeV$, and was also used in previous CMS searches.
For the first time in an HNL search at the LHC, this analysis also uses a data set recorded at $\sqrt{s}=13.6\TeV$, which was collected in 2022 and corresponds to an integrated luminosity of 34.6\fbinv.

The paper is organized as follows.
The CMS detector and the main techniques used for event reconstruction are described in Sections~\ref{sec:CMS_detector} and \ref{sec:event_reconstruction}, respectively.
The considered HNL signal model and the simulation tools are outlined in Section~\ref{sec:signal_simulation}, while Section~\ref{sec:selection} details the event selection and the definition of signal region (SR) bins involved in the statistical analysis.
Section~\ref{sec:background} describes the estimation of residual background contributions using control regions (CRs) in data, and Section~\ref{sec:systematics} discusses the systematic uncertainties.
Section~\ref{sec:results} presents the results of this search and their statistical interpretation, and Section~\ref{sec:summary} provides a summary.
Tabulated results can be found in the HEPData record for this analysis~\cite{hepdata}.

\section{The CMS detector}
\label{sec:CMS_detector}

The CMS detector is composed of several subsystems arranged around one of the collision points of the LHC.
The whole system is divided into three regions: a central barrel region with cylindrical layers and two endcaps to detect particles travelling in the forward direction.
The core feature of the CMS detector is a superconducting solenoid of 6\unit{m} internal diameter, providing a magnetic field of 3.8\unit{T}.
Within the solenoid volume are a silicon pixel and strip tracker, a lead tungstate crystal ECAL, and a brass and scintillator hadron calorimeter (HCAL), each composed of a barrel and two endcap sections.
Forward calorimeters extend the coverage in pseudorapidity ($\eta$) provided by the barrel and endcap detectors.
Muons are detected in gas-ionization chambers covering the range $\abseta<2.4$ and embedded in the steel flux-return yoke outside the solenoid.
In 2016--2018, the muon system comprised three types of detectors: drift tubes (DTs) in the barrel ($\abseta<1.2$), cathode strip chambers (CSCs) in the endcaps ($0.9<\abseta<2.4$), and resistive-plate chambers (RPCs) in both the barrel and the endcaps.
For the 2022 running, an additional set of gas electron multiplier chambers was installed in the forward endcaps, covering
$1.55<\abseta<2.18$~\cite{CMS:2023gfb}.
The muon chambers are assembled into four ``stations'' at increasing distance from the interaction point (IP).
The stations in the barrel are located approximately 4, 5, 6, and 7\unit{m} away from the IP radially, while the stations in the endcaps are located approximately 7, 8, 9.5, and 10.5\unit{m} away from the IP along the beam line axis on both ends of the detector.
Each station provides reconstructed hits in several detection planes, which are combined into track segments, forming the basis of muon reconstruction in the muon system~\cite{CMS_muon_performance_2018}.
A more detailed description of the CMS detector, together with a definition of the coordinate system used and the relevant kinematic variables, can be found in Refs.~\cite{CMS_detector_2008, CMS:2023gfb}.

Events of interest are selected using a two-tiered trigger system.
The first level (L1), composed of custom hardware processors, uses information from the calorimeters and muon detectors to select events at a rate of approximately 100\unit{kHz} within a fixed latency of 4\mus~\cite{CMS_L1_performance}.
The second level, known as the high-level trigger, consists of a farm of processors running a version of the full event reconstruction software optimized for fast processing, and reduces the event rate to a few kHz before data storage~\cite{CMS_trigger_system, CMS:TRG-19-001}.

\section{Event reconstruction}
\label{sec:event_reconstruction}

A particle-flow algorithm~\cite{CMS:2017yfk} aims to reconstruct and identify each individual particle in an event, with an optimized combination of information from the various elements of the CMS detector.
The PV is identified as the vertex corresponding to the hardest scattering in the event and is determined using tracking information alone as described in Section~9.4.1 of Ref.~\cite{CMS-TDR-15-02}.
The energy of photons is obtained from the ECAL measurement.
The energy of electrons is determined from a combination of the electron momentum at the PV as determined by the tracker, the energy of the corresponding ECAL cluster, and the energy sum of all bremsstrahlung photons spatially compatible with originating from the electron track.
The energy of muons is obtained from the curvature of the corresponding track (more details on the reconstruction and identification of muons are provided below).
The energy of charged hadrons is determined from a combination of their momentum measured in the tracker and the matching ECAL and HCAL energy deposits, corrected for the response function of the calorimeters to hadronic showers.
Finally, the energy of neutral hadrons is obtained from the corresponding corrected ECAL and HCAL energies.

For each event, hadronic jets are clustered from these reconstructed particles using an infrared and collinear safe anti-\kt algorithm~\cite{Cacciari:2008gp, Cacciari:2011ma} with a distance parameter of 0.4.
Jet momentum is determined as the vector sum of all particle momenta in the jet, and is found from simulation to be, on average, within 5--10\% of the true momentum over the entire transverse momentum (\pt) spectrum and detector acceptance.
Jet energy corrections are derived from simulation studies so that the average measured energy of jets becomes identical to that of particle-level jets.
In situ measurements of the momentum balance in dijet, photon+jet, {\PZ}+jet, and multijet events are used to determine any residual differences between the jet energy scale in data and in simulation, and appropriate corrections are made~\cite{CMS:2016lmd}.
Additional selection criteria are applied to each jet to remove jets potentially dominated by instrumental effects or reconstruction failures~\cite{CMS:2020ebo}.

Additional \pp interactions within the same or nearby bunch crossings (pileup) can contribute additional tracks and calorimetric energy depositions, increasing the apparent jet momentum.
To mitigate this effect, different strategies are applied for 2016--2018 and 2022 data.
For the former, tracks identified to be originating from pileup vertices are discarded and an offset correction is applied to correct for remaining contributions~\cite{CMS:2020ebo}.
For 2022 data, the pileup-per-particle identification algorithm~\cite{CMS:2020ebo,Bertolini:2014bba} is employed, making use of local shape information, event pileup properties, and tracking information.

Jets originating from the hadronization of \PQb quarks are identified with the \textsc{DeepJet} algorithm~\cite{deepJet_paper, CMS:BTV-16-002}.
We employ a working point with a misidentification rate for light ($\PQu\PQd\PQs\Pg$) jets of 1\%, with a corresponding \PQb quark jet efficiency of 70--80\%, depending on the data-taking year~\cite{btag_WP_Run2, btag_WP_Run3}.

Muon tracks are first separately built in the tracking system (\emph{inner} tracks) and the muon spectrometer (\emph{standalone-muon} tracks).
To accurately reconstruct muons produced close to the IP, such as the prompt muon from the \PW boson decay, we employ the standard CMS muon reconstruction algorithms~\cite{CMS_muon_performance_2012, CMS_muon_performance_2018}, which combine the information from these two subdetectors using two different approaches.
In the \emph{global-muon} reconstruction, standalone-muon tracks are paired with nearby inner tracks using proximity criteria.
When a match is found, a global track fit is performed using the hits from both the tracker and the muon detectors.
The \emph{tracker muon} reconstruction, on the other hand, builds muons by extrapolating the inner tracks to the muon detectors, requiring loose matching with segments in the DT and CSC chambers.
Muon candidates reconstructed by the different algorithms are cross-cleaned and merged into a single muon collection. Global and tracker muons sharing the same inner track, as well as standalone and global muons sharing the same standalone track, are merged into a single candidate; residual overlaps between tracker and standalone muons are removed using segment-based matching. A muon candidate reconstructed with different algorithms is labelled with all the corresponding muon types, \ie global, tracker, or standalone.

The efficiencies of the tracker and global algorithms rapidly decrease for muons originating at distances of more than several cm from the IP.
Hence, to identify the muons from a displaced HNL decay, we use only standalone-muon tracks, \ie rely only on information from the muon detectors.
Standalone-muon reconstruction~\cite{TDR_2006} starts from seeds formed from track segments reconstructed in the DT or CSC chambers.
Starting from each seed, a Kalman-filter track-building procedure propagates the candidate trajectory through the muon system and associates compatible hits in the DT, CSC, and RPC detectors.
The standard standalone-muon reconstruction is optimized for muons produced near the IP and applies a beamspot constraint in the final track fit.
This constraint biases the reconstructed track parameters, reducing the reconstruction efficiency and degrading the momentum resolution for muons originating far from the IP.
For the two muons from the HNL decay, we instead use the displaced-standalone reconstruction.
Its seeds are constructed using a procedure derived from cosmic ray muon reconstruction, without requiring the trajectory to point towards the IP, and the final track fit is performed without a beamspot constraint.
The resulting tracks are reconstructed exclusively from measurements in the muon system and are referred to as displaced standalone (DSA) muons. In the following, the term standalone (STA) muons is reserved for tracks reconstructed with the standard standalone-muon algorithm. For the 2016--2018 selection, matching STA muons provide the timing information used to reject muons from neighbouring bunch crossings, as described in Section~\ref{subsec:muon_selection}.

For the 2022 data, the displaced-muon collection additionally contains displaced-tracker (dTK) and displaced-global (dGLB) muons. The dTK muons are formed from inner-tracker tracks reconstructed using muon system information and matched to compatible muon segments. The dGLB muons are reconstructed by propagating DSA tracks toward the tracker and associating them with compatible inner tracks. Both types therefore contain tracker information.
Similarly to the standard muon collection, the candidates reconstructed by the DSA, dTK, and dGLB algorithms are
cross-cleaned and merged into a single displaced-muon collection. If the same muon is reconstructed by more than one of these algorithms, the merged object retains the corresponding muon type labels.

\section{Signal model and simulation}
\label{sec:signal_simulation}

We employ the \MGvATNLO v2.6.5 event generator~\cite{Alwall:2014hca} to generate Monte Carlo (MC) event samples for the HNL production process simulated at leading order (LO), using the model described in Ref.~\cite{HeavyN_model_theory} and provided in the UFO format by Refs.~\cite{HeavyN_model_UFO_Alva,HeavyN_model_UFO_Degrande}.
The NNPDF3.1 parton distribution functions~\cite{NNPDF:2017mvq} at next-to-next-to-LO (NNLO) accuracy in quantum chromodynamics (QCD) are used for simulating the hard process, and the \PYTHIA v8.240 (v8.306)~\cite{Sjostrand:2014zea, Bierlich:2022pfr} event generator with the CP5 tune~\cite{CMS-GEN-17-001} is used to simulate the parton shower for samples simulated at 13 (13.6)\TeV.
Following Ref.~\cite{CMS-EXO-20-009}, the production cross section is scaled to NNLO QCD precision using the ratio of the cross section for \PW boson production calculated at NNLO with the \FEWZ v3.1 program~\cite{Melnikov:2006kv,Gavin:2010az,Gavin:2012sy,Li:2012wna} and the one from LO \PW boson samples produced with the exact same settings as the signal samples.
Simulated minimum bias events are superimposed onto the hard collisions in the simulated signal samples to reproduce the correct pileup distributions.
All generated events are then processed using a simulation of the CMS detector based on the \GEANTfour software~\cite{Geant4}.

Signal events are generated for the production of one HNL from a leptonic \PW boson decay and the subsequent HNL decay into two charged leptons and a neutrino, including HNL decays mediated both by \PWst and \PZst bosons.
We generate samples corresponding to \mN values in the range of 1--6\GeV, where the HNLs are assumed to couple exclusively to the second-generation neutrino (\PGnGm).
For each mass hypothesis, the value of the coupling strength \VNm used in the simulation is chosen to maximize the fraction of generated events within the analysis acceptance, which approximately corresponds to a transverse HNL decay length between 50 and 400\unit{cm}, yielding values of \VNmsq between $10^{-2}$ and $10^{-7}$ for different \mN values.

Since the values of \mN and \VNmsq affect not only the HNL production cross section but also its lifetime, they have an impact on the analysis acceptance and on the reconstruction efficiency.
Hence, a simple cross section rescaling of the signal predictions based on the different coupling strength does not yield the correct estimates of the expected signal rates at different \VNmsq values.
To achieve this, a per-event reweighting of each simulated sample is performed so that the HNL proper lifetime distribution (before the parton-shower and detector simulation) matches the expected proper lifetime distribution for the target \VNmsq value.

\section{Event selection and categorization}
\label{sec:selection}

Signal event candidates are expected to contain one prompt muon from the \PW boson decay, denoted as \muzero, and a pair of displaced muons with opposite electric charge from the HNL decay, denoted as \muone and \mutwo and numbered in order of decreasing \pt.
The SM background candidates generally originate from processes producing at least one prompt muon, such as Drell--Yan, {\PW}+jets, \ttbar, and diboson production. These events can mimic the signal when the prompt muon is accompanied by an additional pair of muons, \eg from a pileup \PJGy meson decay or semileptonic decays of \PQb hadrons.
The event selection is optimized using simulated HNL signal samples and background-enriched CRs in data, while the residual background yields after the full selection are determined with the method based on CRs in data, as described in Section~\ref{sec:background}.

For a given event, all possible pairs of DSA muons are constructed.
For each pair, a fit of the two tracks to a common SV is performed with a fitting algorithm based on the Kalman filter approach~\cite{Fruhwirth:1987fm}.
Subsequently, only events with one prompt muon and at least one pair of DSA muons with a converging SV fit are retained.
In addition, a DSA muon pair is discarded if any of the two tracks is matched to a muon with tracker hits, making our identification criteria orthogonal to the ones used in Ref.~\cite{CMS-EXO-20-009}.
The set of identification criteria applied to the reconstructed muons are detailed in Section~\ref{subsec:muon_selection}.

Given the low HNL masses $\mN\lesssim5\GeV$ targeted by this search, the HNLs produced in the decay of a \PW boson are generally highly Lorentz boosted.
Consequently, \muone and \mutwo from the HNL decay are produced in a collimated cone whose direction is aligned with the momentum vector of the HNL.
Furthermore, in the absence of large hadronic activity, the \PW boson is typically produced with low \pt~\cite{CMS:SMP-18-012}, resulting in large azimuthal angles between \muzero and the HNL, and consequently between \muzero and both \muone and \mutwo.
All these features of the signal topology are exploited in the analysis to design the kinematic selection criteria that define the SR, detailed in Section~\ref{subsec:kinematic_selection}.

To further enhance the analysis sensitivity, the selected events are divided into four SR bins defined by two variables that provide significant signal-to-background discrimination.
The first variable is the displaced dimuon transverse momentum $\ptmumu=\abs{\ptvec(\muone)+\ptvec(\mutwo)}$, where each \ptvec is the muon momentum projected onto the plane perpendicular to the beam axis.
This variable measures the \pt of the visible dimuon system, and its value is typically higher in signal than in background events because of the HNL boost.
The second variable, \Lxysig, is the significance of the reconstructed SV displacement with respect to the PV, where \Lxy is measured as the distance in the $x$--$y$ plane between the SV and PV, and $\sigma_{\Lxy}$ is its uncertainty.
This variable measures the significance of the SV displacement from the PV: it tends to be large for genuine HNL decays, while background dimuon pairs are predominantly prompt or only mildly displaced.
More details on the SR bins and their optimization are provided in Section~\ref{subsec:binning_optimisation}.

\subsection{Muon selection}
\label{subsec:muon_selection}

We analyze events recorded by single-muon triggers~\cite{CMS:MUO-19-001}, which require the presence of at least one isolated muon at the high-level trigger with $\pt>27$ (24)\GeV in 2017 (all other years).
We then select events with exactly one prompt muon with $\pt>29$ (26)\GeV in 2017 (all other years) that meets the set of ``tight'' muon identification (ID) criteria defined in Ref.~\cite{CMS_muon_performance_2018}, which are based on the track quality and the compatibility between the inner track and the measurements in the muon system.
The relative isolation variable \Irel is defined as the scalar sum of the \pt of charged hadrons originating from the PV, neutral hadrons, and photons within a cone around the muon, divided by the muon \pt, where the cone size is defined by $\DR=\sqrt{\smash[b]{(\Deta)^2+(\Dphi)^2}}<0.4$ with \Deta and \Dphi being the $\eta$ and azimuthal angle differences between the particles and the muon, respectively.
We require that the prompt muon is isolated from any hadronic activity by imposing $\Irel<0.15$.
Additionally, it must be associated with the PV by having a transverse impact parameter of less than 50\mum, and being geometrically
matched to a trigger-level muon by requiring $\DR<0.1$.
The prompt muon passing these selection criteria is identified as \muzero.

Each DSA muon is required to have $\pt>5\GeV$, a relative uncertainty in the measured \pt smaller than 100\%, a number of associated hits in the muon system greater than 15, and a reduced \chisq value of the track fit smaller than 2.5.
In addition, the DSA muon must not have any associated tracker information. This requirement suppresses background from prompt SM sources and ensures no overlap with the SR selections of the analysis in Ref.~\cite{CMS-EXO-20-009}.
The reconstructed muon time must also be within $10\unit{ns}$ of the primary interaction time, in order to reject muons originating from neighbouring bunch crossings, for which the corresponding tracker hits are not recorded in the event.
These two criteria are further detailed in the following paragraphs.

The reconstructed data format in 2016--2018 did not contain information about the time of the muon hits for the DSA muons.
Since the time selection is crucial to remove background from muons produced in neighbouring bunch crossings, we use the time information from matching STA muons.
Each DSA muon in the 2016--2018 data and simulation is therefore required to have a unique STA-muon match with $\DR(\text{DSA},\text{STA})<0.4$, where \DR is the separation between the DSA and STA track directions in the $\eta$--$\phi$ plane.
Two timing estimates associated with the matched STA muon are used. The first is reconstructed using measurements from the full muon system, while the second uses only RPC measurements and exploits the nanosecond-level RPC time resolution for bunch crossing assignment.
Both estimates are required to be within 10\unit{ns} from the primary interaction time.
The DSA-STA matching requirement effectively limits the 13\TeV results to the reconstruction efficiency of STA muons.
However, using DSA muons instead of STA muons allows us to benefit from their higher momentum resolution, which improves the efficiency of the kinematic selection criteria described in Section~\ref{subsec:kinematic_selection}.
For the 2022 data and simulation, the time information was added to the analysis data format, and thus the time selection criteria can be directly applied to the DSA muons.

In 2016--2018 data and simulation, to reject DSA muons with associated tracker activity, we evaluate whether the matched STA muon is also reconstructed with the global or tracker muon algorithm.
Additionally, we reject DSA muons that are matched to a tracker or global muon with $\pt>3\GeV$ that passes the set of ``medium'' muon ID criteria defined in Ref.~\cite{CMS_muon_performance_2018}, where the matching is defined by $\DR(\text{DSA},\PGm)<0.4$.
In 2022 data and simulation, we apply a similar veto by first requiring the DSA muons not to be also reconstructed as dTK or dGLB muons, and then by requiring no global or tracker muon passing the same criteria as for 2016--2018 to be matched to the DSA muon within a cone of $\DR<0.4$.

Since DSA muons from background sources are typically less isolated than those from the signal process, we apply further requirements on the isolation of the DSA muons.
In 2016--2018 data and simulation, events are rejected if either of the two DSA muons has a charged isolation \Ich, calculated as the sum of the \pt of all charged hadrons in a cone of $\DR<0.4$ around the muon, larger than 5\GeV.
For 2022 data and simulation, we use the relative charged isolation \Irelch, defined as the scalar sum of the \pt of charged hadrons in a cone of $\DR<0.3$ around the muon, divided by the muon \pt.
We require $\Irelch<0.5$ for both DSA muons.

Finally, we require the presence of at least two DSA muons with opposite charges, and identify them as \muone and \mutwo.
In signal simulation, this requirement removes less than 3\% of selected signal events in the considered mass range, while it reduces the background by almost a factor of two.
If more than one pairing of DSA muons with opposite charge is found, the pair with the highest \ptmumu is selected.

In 2016--2018, the DSA muon selection has an efficiency of 17--20\% for
simulated signal samples with masses between 1 and 5\GeV, while
suppressing the background by more than five orders of magnitude.
This efficiency includes the reconstruction of two DSA tracks and the SV, the DSA identification criteria, the matching with STA muons, the vetoes on tracker and out-of-time muons, and the final charge requirement.
In 2022, a relative increase in signal efficiency of 10--15\% is observed because the timing information is available directly for DSA muons, removing the STA matching requirement used for 2016--2018.
These values are estimated using simulated events that pass the trigger and prompt-muon selections, and in which the HNL decay produces two muons within the CMS acceptance, \ie with $\pt>3\GeV$ and $\abseta<2.4$.
As mentioned previously, signal samples are generated, for each mass point, with a value of \VNmsq that maximizes the number of events with transverse decay lengths \Lxy between 50 and 400\cm, resulting in similar \Lxy distributions in all cases.
Since the analysis acceptance strongly depends on \Lxy, and \Lxy itself depends on \VNmsq, the quoted efficiency varies with the coupling strength.
For example, values of \VNmsq that are twice as large as those used in the simulated samples lead to a relative loss in signal efficiency of up to 20\%.

\subsection{Kinematic selection}
\label{subsec:kinematic_selection}

Additional kinematic selections are introduced to further suppress the remaining background contributions by exploiting the signal topology.
To reduce background from processes with top quarks, events are discarded that have at least one \PQb-tagged jet with $\pt>25\GeV$ and $\abs{\eta}<2.5$.
To suppress $\PZ\to\MM$ contributions, events are rejected if the invariant mass \mZveto of the prompt muon \muzero combined with the highest \pt muon \muprime in the event (excluding \muzero) exceeds 80\GeV.
For this selection, \muprime is chosen among the remaining muons in the event passing the ``loose'' muon ID criteria defined in Ref.~\cite{CMS_muon_performance_2018}. This choice effectively excludes STA muons, whose typically poor \pt resolution in signal events could cause them to satisfy the veto requirement and thus reduce the signal efficiency.

The quality of the SV fit is estimated through a $p$-value \pSV, \ie a probability based on the \chisq value of the vertex fit.
A requirement of $\pSV>0.001$ is applied to remove DSA muon pairs that are not consistent with originating from the same vertex.
The fit quality is also reflected in the uncertainty in the SV position, which enters the calculation of \Lxysig described in Section~\ref{subsec:binning_optimisation} to discriminate between signal and background contributions.

Given the large boost of the HNL, \muone and \mutwo are expected to be produced within a small angle in the $\eta$--$\phi$ plane.
In addition, since the undetected neutrino from the HNL decay is also produced within the same small angular cone, the combined momentum of the dimuon system is generally aligned with the direction of the HNL momentum.
Hence, the requirements $\DRmumu<1$ and $\costheta>0.9$ are applied, where $\theta$ is the angle between the vector sum of the momenta of \muone and \mutwo and the vector from the PV to the SV, which defines the reconstructed direction of the HNL momentum.
Another consequence of the large HNL boost is that signal events generally have higher \ptmumu values than background contributions, and we thus require $\ptmumu>10\GeV$.
This variable is also used in the event categorization described in Section~\ref{subsec:binning_optimisation}.

Finally, the three final-state muons must be consistent with the decay topology of an on-shell \PW boson.
The invariant mass of the three muons, \mmumumu, is constrained by the \PW boson mass.
Given the relatively low momentum carried by the neutrino, the distribution of this mass peaks just below the \PW boson mass, with a steep decrease above 80\GeV and a larger tail in the low-mass range.
Therefore, we require events to have $40<\mmumumu<90\GeV$, to reduce background contributions, while maintaining a high signal efficiency.
Additionally, \muzero typically recoils at large angles in the transverse plane from the HNL, so we require the azimuthal angle \Dphimumu between \muzero and both \muone and \mutwo to be greater than 1.5.

\begin{table}[!ht]
\centering
\topcaption{Summary of event selection criteria defining the SR.}
\label{table:selection_criteria}
\renewcommand{\arraystretch}{1.4}
\begin{tabular}{lc}
    \hline
    Description & Criterion \\
    \hline
    \PZ boson veto & $\mZveto<80\GeV$ \\
    \PQb jet veto & $N_{\text{b-tagged jets}}=0$ \\
    SV fit quality & $\pSV>0.001$ \\
    \PW boson decay & $\Dphimumu>1.5$ \\
    HNL boost & $\DRmumu<1.0$ \\
    & $\ptmumu>10\GeV$ \\
    & $\costheta>0.9$ \\
    \PW boson mass & $40<\mmumumu<90\GeV$ \\
    \hline
\end{tabular}
\end{table}

\begin{figure}[!p]
\centering
\includegraphics[width=0.47\textwidth]{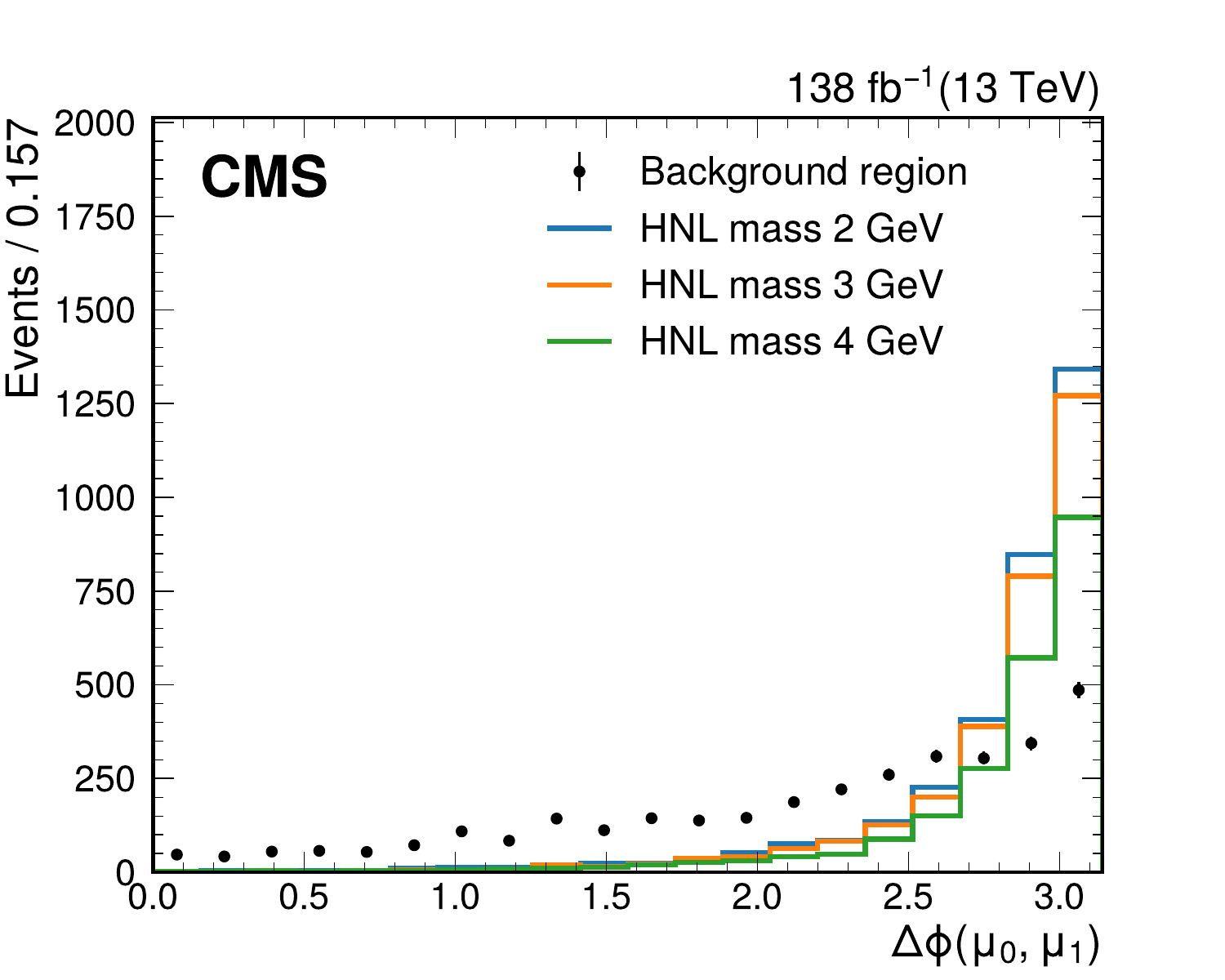}%
\hspace*{0.05\textwidth}%
\includegraphics[width=0.47\textwidth]{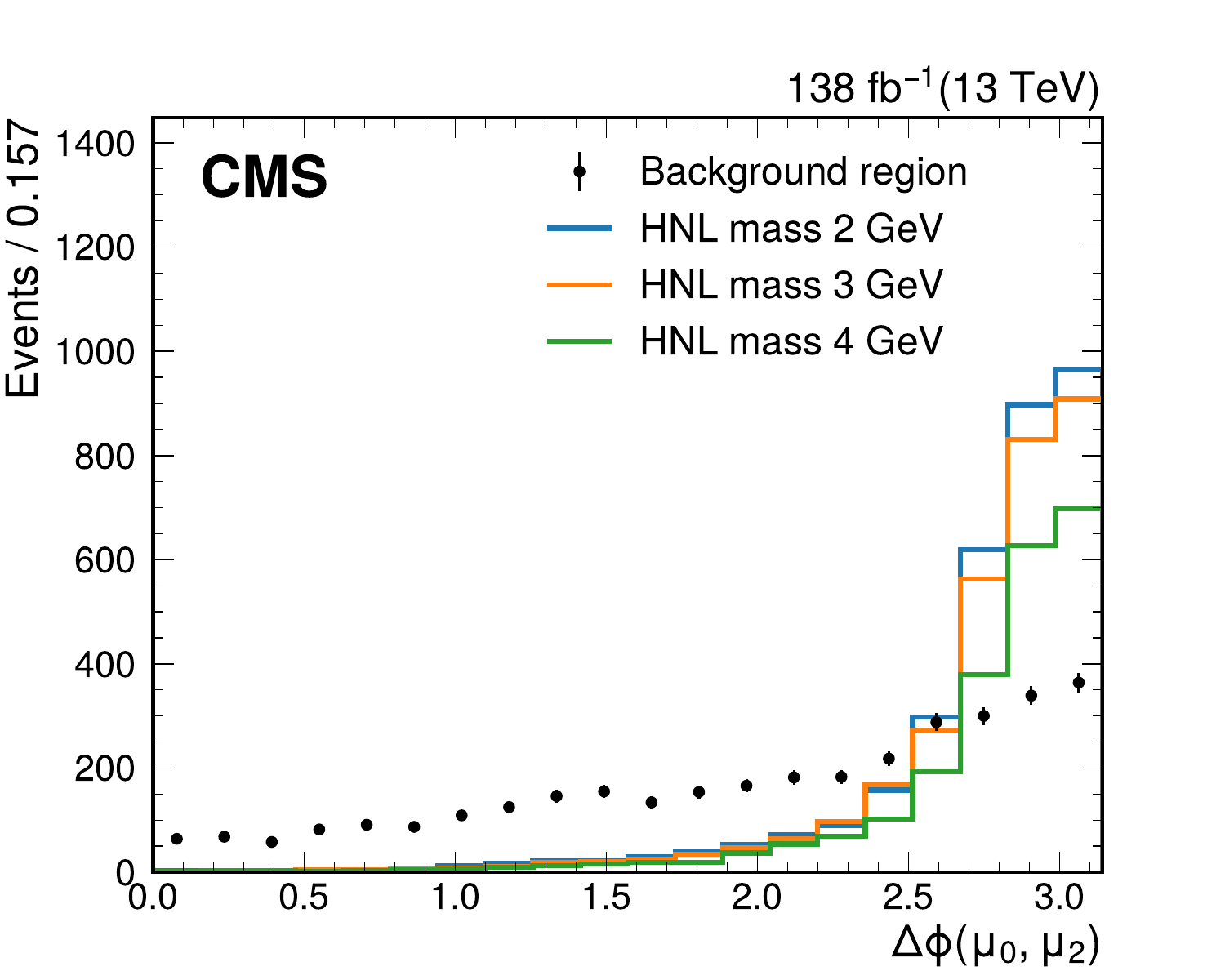} \\
\includegraphics[width=0.47\textwidth]{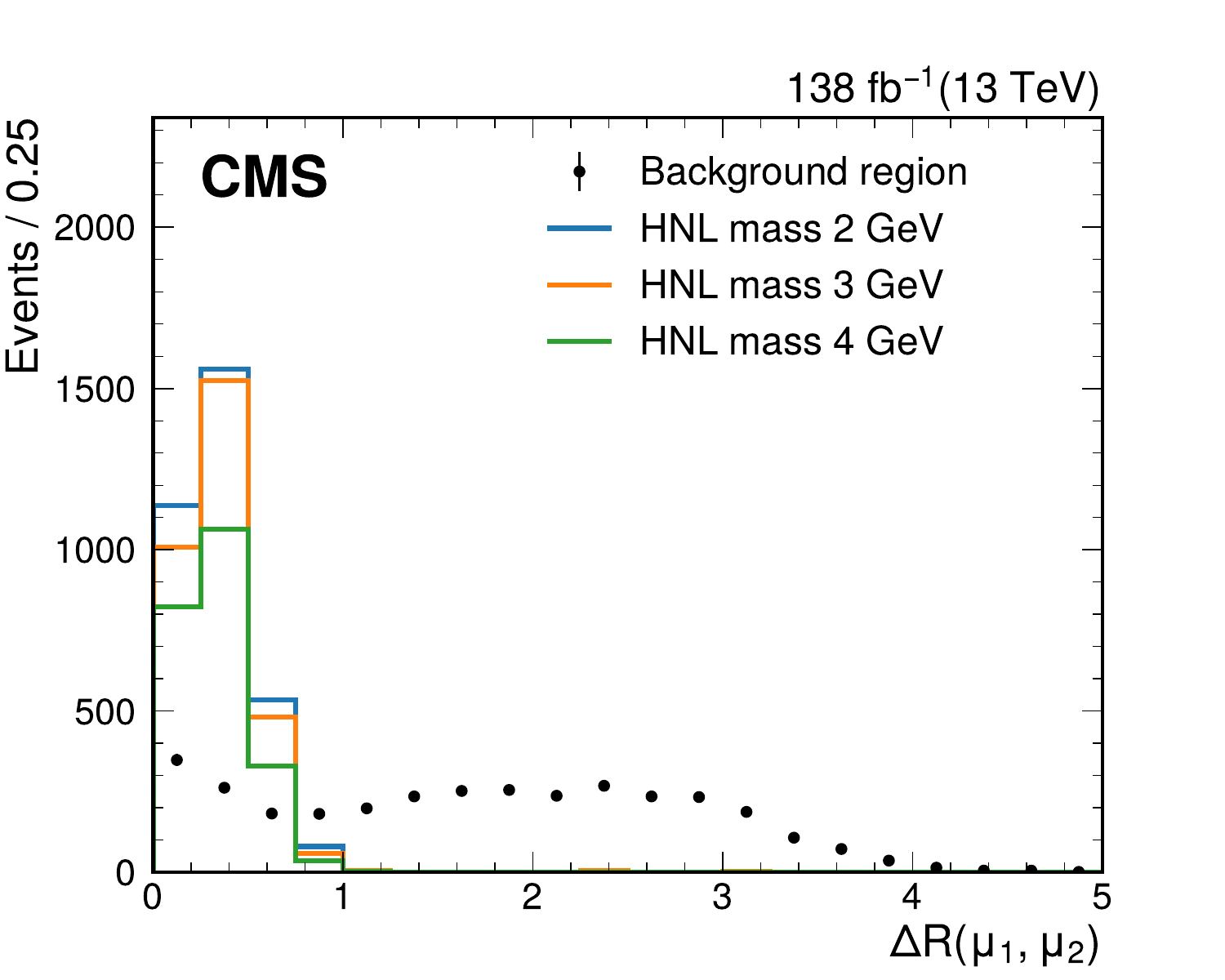}%
\hspace*{0.05\textwidth}%
\includegraphics[width=0.47\textwidth]{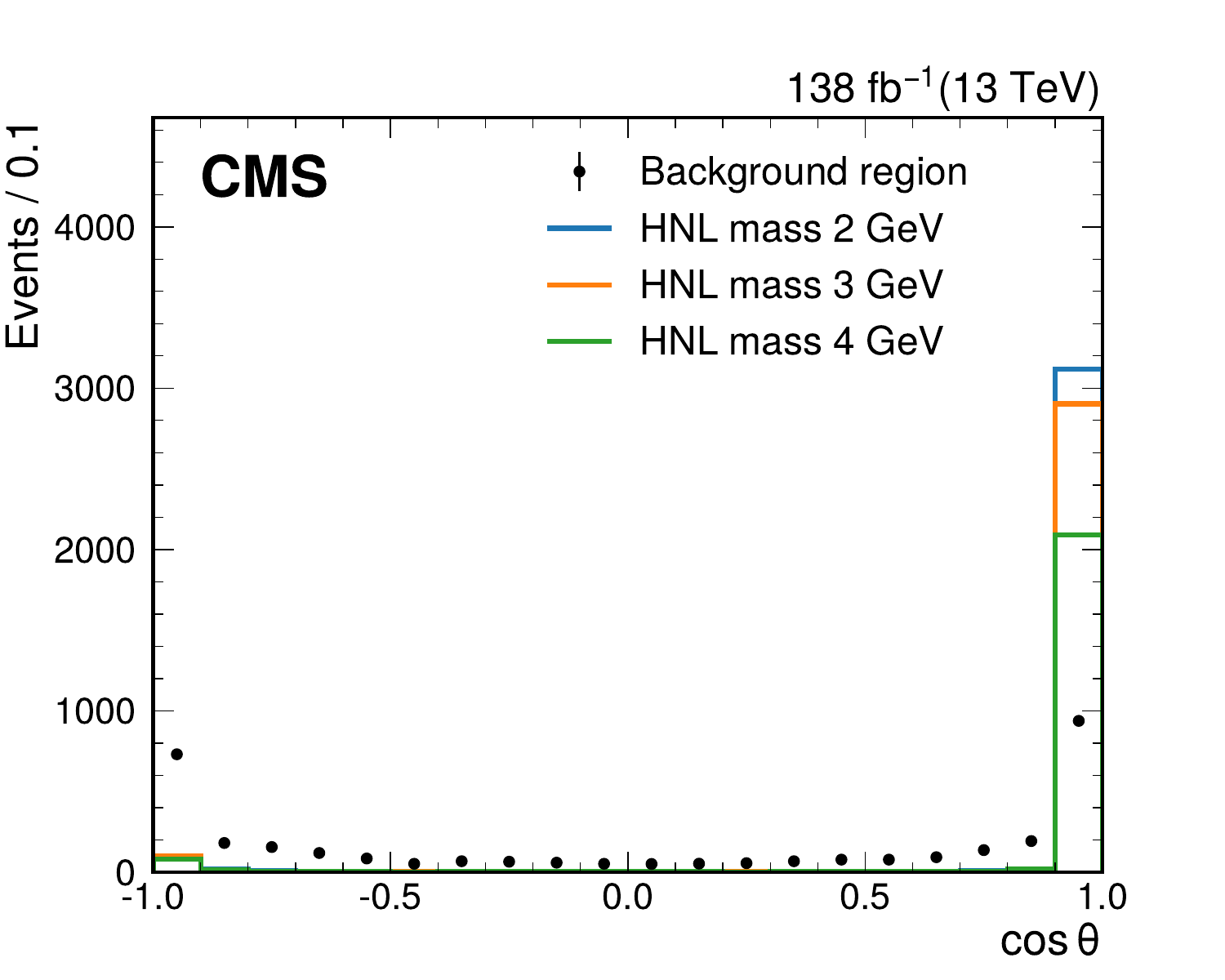} \\
\includegraphics[width=0.47\textwidth]{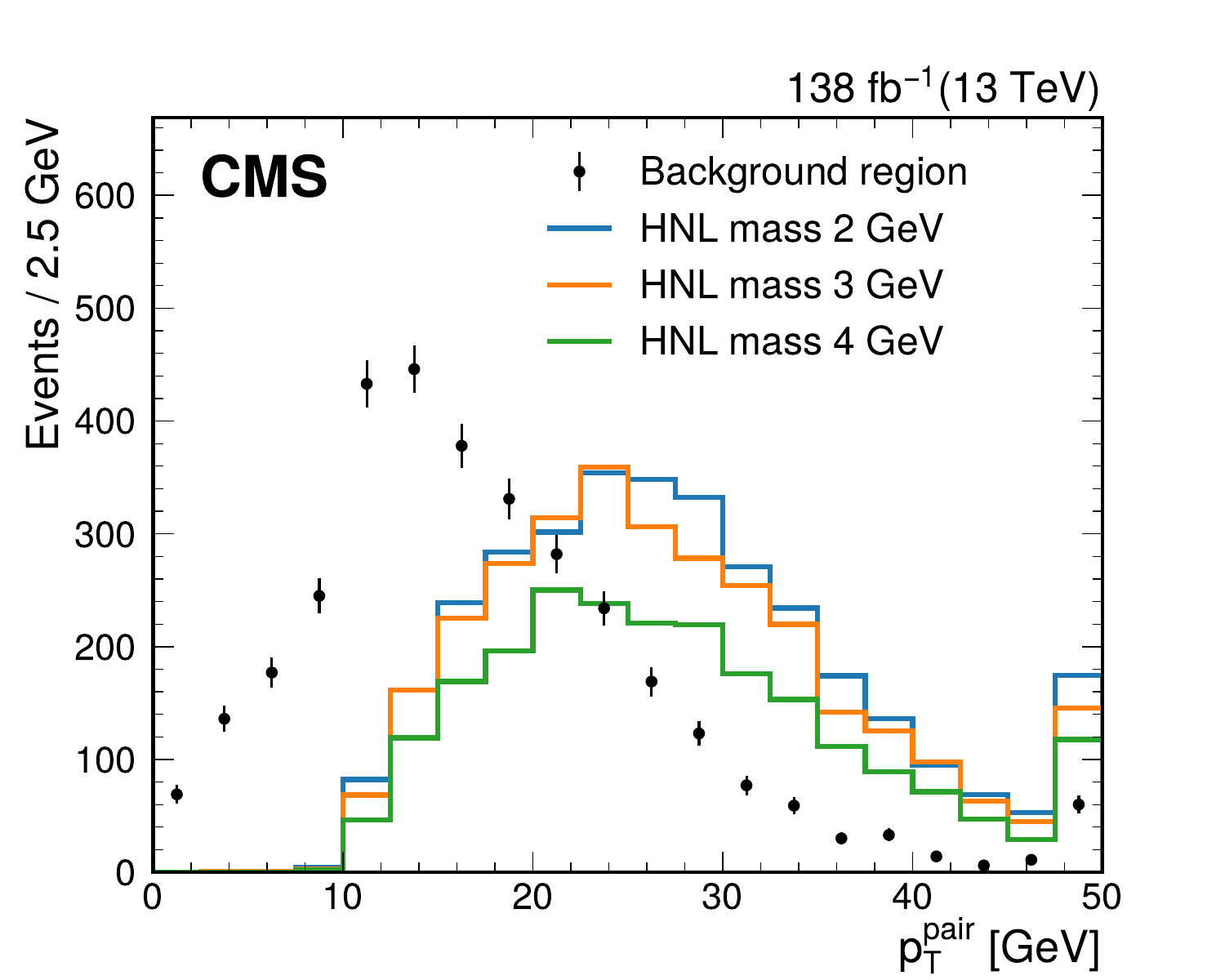}%
\hspace*{0.05\textwidth}%
\includegraphics[width=0.47\textwidth]{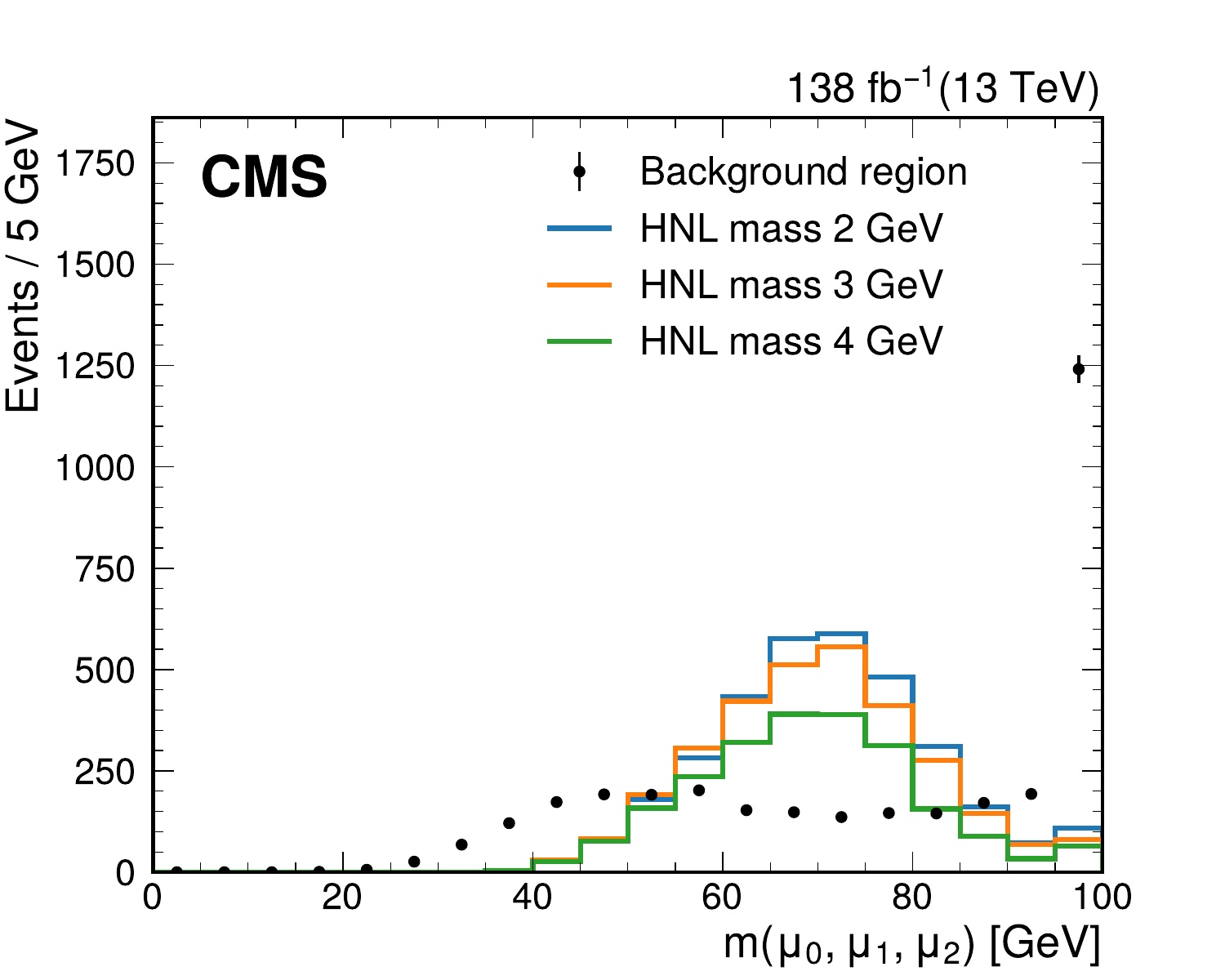}
\caption{%
    Distributions of $\Dphi(\muzero,\muone)$ (upper left), $\Dphi(\muzero,\mutwo)$ (upper right), \DRmumu (middle left), \costheta (middle right), \ptmumu (lower left), and \mmumumu (lower right), shown for HNL hypotheses with three different \mN values in the SR.
    The distributions are taken after the muon selection and compared to the background distribution from the CR in data obtained by inverting the tracker veto on \muone (``background region'').
    The HNL predictions are scaled with a single normalization factor such that the integral of the HNL prediction with $\mN=2\GeV$ matches the integral of the data distribution, with relative normalization differences between different HNL hypotheses retained.
    Overflow contributions are included in the last bin of the histograms, where appropriate.
}
\label{fig:dist_kin_selection}
\end{figure}

The phase space defined by these selection criteria, summarized in Table~\ref{table:selection_criteria}, is referred to as the SR.
The cumulative efficiency of all the kinematic selection criteria estimated with signal MC samples having \mN between 1 and 5\GeV is within 80--83\%, while the background is reduced by more than a factor of 20.
Distributions for several of the observables used in the event selection are shown in Fig.~\ref{fig:dist_kin_selection}.
Signal distributions are obtained from the MC samples for different HNL scenarios.
Since the background MC samples were found to inadequately describe the dominant background contributions arising from instrumental inefficiencies, distributions of background after the muon selection are obtained from a CR in data where the tracker veto on \muone is inverted, as defined in Section~\ref{sec:background}.

\subsection{Optimization of signal regions}
\label{subsec:binning_optimisation}

To maximize the sensitivity to the HNL signal, we bin the SR in \ptmumu and \Lxysig.
The distribution of \ptmumu for the HNL signal peaks around 25\GeV, while background contributions have typically lower values and fall off above $\sim$15\GeV.
The \Lxysig distribution presents a steeply falling shape for background processes, which primarily arise from misreconstructed prompt or mildly displaced signatures, whereas larger \Lxy values are characteristic of genuine displaced HNL decays.
Besides presenting distinct shapes for signal and background, as well as a low correlation with the transfer factor (\TF) used to derive the background prediction, the two variables are also largely statistically independent of each other, allowing us to exploit the discriminating power of both simultaneously.

\begin{figure}[!ht]
\centering
\includegraphics[width=0.5\textwidth]{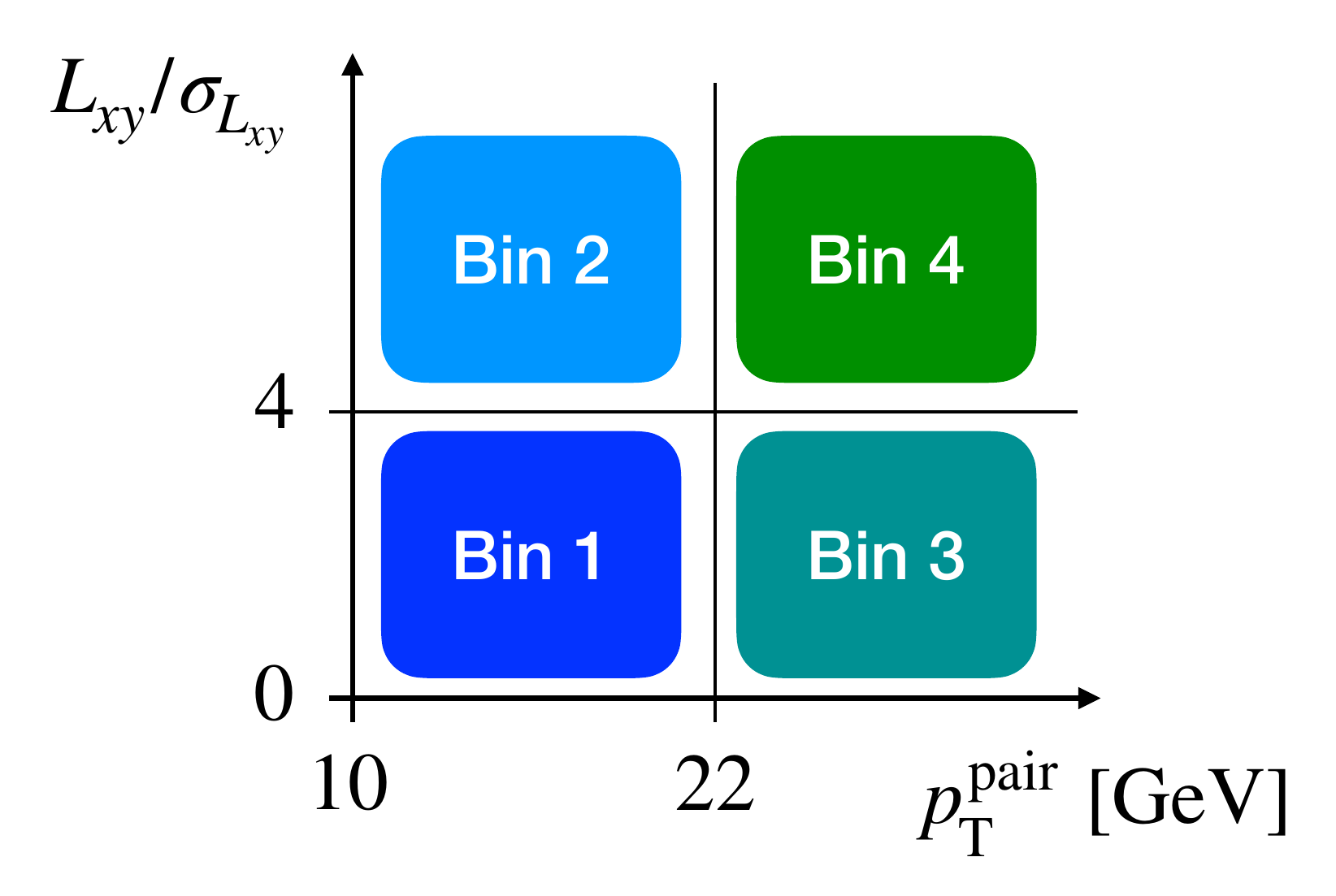}
\caption{Definition of the SR bins.}
\label{fig:signal_bins}
\end{figure}

To optimize the binning in these two variables, we use the smallest \VNmsq value excluded at 95\% confidence level (\CL) by Ref.~\cite{CMS-EXO-20-009} in the scenario with a Majorana HNL and consider the corresponding signal yields and distributions.
Representative background distributions are taken from a CR in data where the tracker veto on \muone is inverted, \ie where \muone is matched with an in-time muon passing the medium muon ID, scaled down by a constant \TF, as described in Section~\ref{sec:background}.
The optimal binning definition is determined by maximizing the expected discovery significance for the HNL signal separately for each \mN hypothesis, but we find similar results in all considered cases.
As a result, we define a single set of four bins that are used in the search for all \mN hypotheses.
The SR bins are defined by thresholds of $\Lxysig=4$ and $\ptmumu=22\GeV$, as shown in Fig.~\ref{fig:signal_bins}.
The highest contribution to the discovery significance for the considered HNL hypotheses is provided by \mbox{bin 4} with $\ptmumu>22\GeV$ and $\Lxysig>4$.

\section{Background estimation}
\label{sec:background}

The background from SM processes that contribute to the SR originates from non-displaced sources of dimuons produced in association with a muon satisfying the prompt-muon selection.
These dimuons, comprising both resonant and non-resonant contributions, are misreconstructed as  displaced muon pairs because of inefficiencies of the tracking system or the reconstruction process.
Pairs of resonant muons mainly originate from pileup \PJGy meson decays or cascade decays of \PQb hadrons, while the primary source of non-resonant muon pairs is inclusive \PQb hadron production, where two semileptonic \PQb hadron decays produce independent muons.
For these background contributions, MC event samples are not expected to provide reliable predictions as the simulation may not accurately reproduce instrumental effects, such as detection inefficiencies or reconstruction mistakes.

Thus, we predict the background from control samples in data using the so-called ``ABCD method'', which relies on two statistically independent conditions referred to as \emph{tracker veto on \muone} and \emph{reduced signal selection}.
An event passes the tracker veto on \muone if this muon does not overlap with a global or tracker muon, according to the separate criteria defined for the 2016--2018 and 2022 analyses in Section~\ref{subsec:muon_selection}.
The reduced signal selection consists of the following subset of the
event selection requirements introduced in Section~\ref{subsec:kinematic_selection}:
\begin{itemize}
    \item $\ptmumu>10\GeV$,
    \item $40<\mmumumu<90\GeV$,
    \item $\Dphimumu>1.5$.
\end{itemize}
An event passes the reduced signal selection if all three requirements are satisfied, and fails it if at least one requirement is failed.
This subset of requirements was optimized to minimize correlations with the tracker veto, retain a background composition similar to that in the SR, and provide sufficient event yields in the CRs.

The two conditions define four regions, as summarized in Table~\ref{tab:abcd_defintion}.
Region~A, in which both conditions are satisfied, corresponds to the SR, while regions B, C, and D are dominated by background.
Under the assumption that the two above conditions are statistically independent, the background yield in region A is estimated as
\begin{equation}\label{eq:bkg_estim}
    \numA=\numB\frac{\numC}{\numD}\equiv\numB\,\TF,
\end{equation}
where the ratio of the yields in regions C and D is defined as the \TF.

\begin{table}[ht!]
\centering
\topcaption{Definition of the regions used in the ABCD method.}
\label{tab:abcd_defintion}
\renewcommand{\arraystretch}{1.2}
\begin{tabular}{lcc}
    \hline
    Region & Tracker veto & Reduced signal selection \\
    \hline
    A & Pass & Pass \\
    B & Fail & Pass \\
    C & Pass & Fail \\
    D & Fail & Fail \\
    \hline
\end{tabular}
\end{table}

The signal contamination in regions B, C, and D was estimated for several HNL hypotheses with masses between 1 and 5\GeV and \VNmsq corresponding to the smallest excluded values at 95\% CL in Ref.~\cite{CMS-EXO-20-009}.
In all cases, the signal contamination was found to be negligible.
In regions B and D, the signal yields amount to less than 1\% of the observed data yield.
In region C, where the reduced signal selection is inverted, the contamination is at most 6\%, which is also negligible compared to the \TF uncertainty derived from this region.

To measure the \TF in data, we use an event selection with a relaxed isolation criterion applied to the DSA muons in order to reduce the statistical uncertainty in the \TF.
Specifically, the \Ich requirement is not applied to DSA muons in 2016--2018, and a looser requirement of $\Irelch<5$ is applied in 2022 data and simulation.
The yields observed for the four bins combined in regions B, C, and D together with the \TF evaluated from these yields are reported in Table~\ref{tab:abcd_yields}, separately for the 2016--2018 and 2022 data sets.
The \TF is used as a constant scale parameter applied on the observed rates in region B for the background estimation after the full SR selection.
The robustness of the ABCD method is assessed by comparing the background prediction with the observed number of events in two distinct validation regions (VRs), as described in the following.

\begin{table}[!ht]
\centering
\topcaption{%
    Observed yields in the regions B, C, and D for the selection with relaxed DSA isolation requirements, and corresponding transfer factor obtained as the ratio of events in the regions C and D.
}
\label{tab:abcd_yields}
\renewcommand{\arraystretch}{1.2}
\begin{tabular}{lcccc}
    \hline
    Signal region with & \multicolumn{3}{c}{Observed} & Transfer factor\\[-3pt]
    relaxed DSA isolation & \numB & \numC & \numD & \numC/\numD \\
    \hline
    2016--2018 & 198 & 16 & 132 & $0.12 \pm 0.03$ \\
    2022  & 29  & 5  & 25  & $0.2 \pm 0.1$ \\
    \hline
\end{tabular}
\end{table}

\subsection{Validation with low-\texorpdfstring{\pt}{pT} muons}

We define a low-\pt validation region by requiring $3<\pt(\muonetwo)<5\GeV$ for each DSA muon and removing the requirement of $\ptmumu>10\GeV$.
Because of the high HNL boost, signal events will generally have higher \pt values of the DSA muons and thus this validation region has a very low signal contamination.
Instead, a significant yield of background events with soft muons misidentified as highly displaced is expected.
The yields observed for the four bins combined in regions A, B, C, and D, using relaxed isolation criteria identical to those used for Table~\ref{tab:abcd_yields}, together with the predicted yield in region A, are reported in Table~\ref{tab:abcd_yields_lowPt}.
Good agreement between the observed and predicted yield in region A is found.

\begin{table}[!ht]
\centering
\topcaption{%
    Observed yields in the regions A, B, C, and D for the low-\pt validation region with relaxed DSA isolation requirements, and the yield in region A predicted using the ABCD method.
}
\label{tab:abcd_yields_lowPt}
\renewcommand{\arraystretch}{1.2}
\begin{tabular}{lccccc}
    \hline
    Low-\pt & \multicolumn{4}{c}{Observed} & Predicted \\[-3pt]
    validation region & \numA & \numB & \numC & \numD & \numA \\
    \hline
    2016--2018 & 22 & 116 & 29 & 180 & $19 \pm 4$ \\
    2022  & 6  & 24  & 9  & 58  & $3.7 \pm 1.5$ \\
    \hline
\end{tabular}
\end{table}

\begin{figure}[!tp]
\centering
\includegraphics[width=0.47\textwidth]{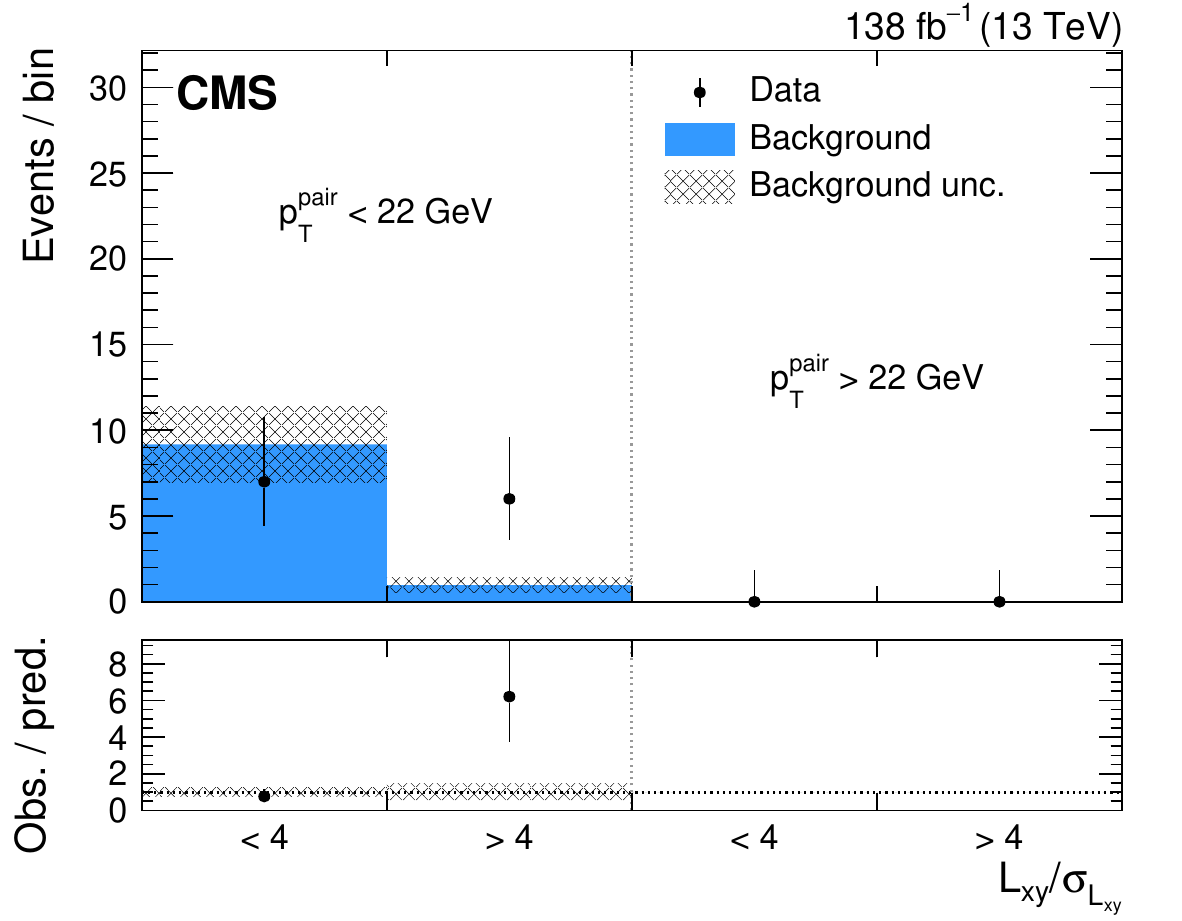}%
\hspace*{0.05\textwidth}%
\includegraphics[width=0.47\textwidth]{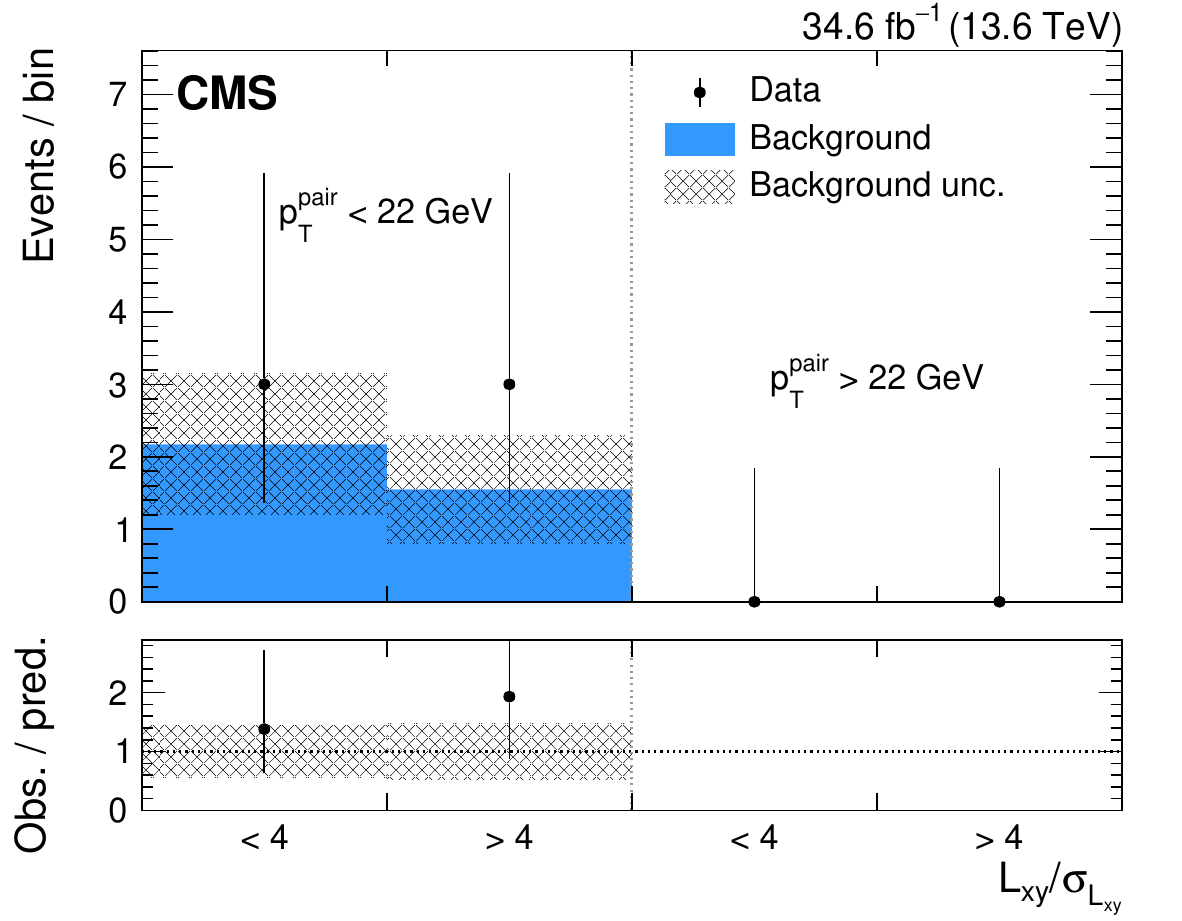}
\caption{%
    Observed (black points) and predicted (blue histograms) yields in the low-\pt validation region in 2016--2018 (left) and in 2022 with relaxed DSA isolation requirements (right).
    The hatched bands represent the statistical uncertainty in the background prediction.
    The lower panels show the ratio between the observed and predicted yields in each bin.
    Both predicted and observed yields are exactly zero in the last two bins by construction of the low-\pt validation region.
}
\label{fig:closure_lowPt}
\end{figure}

Additionally, we consider this validation separately for the four bins of \ptmumu and \Lxysig.
In Fig.~\ref{fig:closure_lowPt}, the comparison of the observed yields in region A of the low-\pt validation region with the predicted yields as evaluated with the ABCD method is shown.
The 13\TeV data are shown with the nominal DSA isolation requirements, whereas the 2022 data are shown with relaxed DSA isolation requirements to increase the statistical power of the validation.
The observed yields are compatible with the prediction except for \mbox{bin 2} in the 2016--2018 data.
This excess of data over the ABCD prediction can be explained by photons converting into muons: low-\pt photons produced in hadronic jets can convert into two muons when interacting with the detector material, thereby generating genuinely displaced muon pairs that are not accounted for by the ABCD method.
Since the \pt distribution of photons from jets is steeply falling, the contribution is expected to be negligible at high \pt, \ie in the SR.

\begin{figure}[!htp]
\centering
\includegraphics[width=0.47\textwidth]{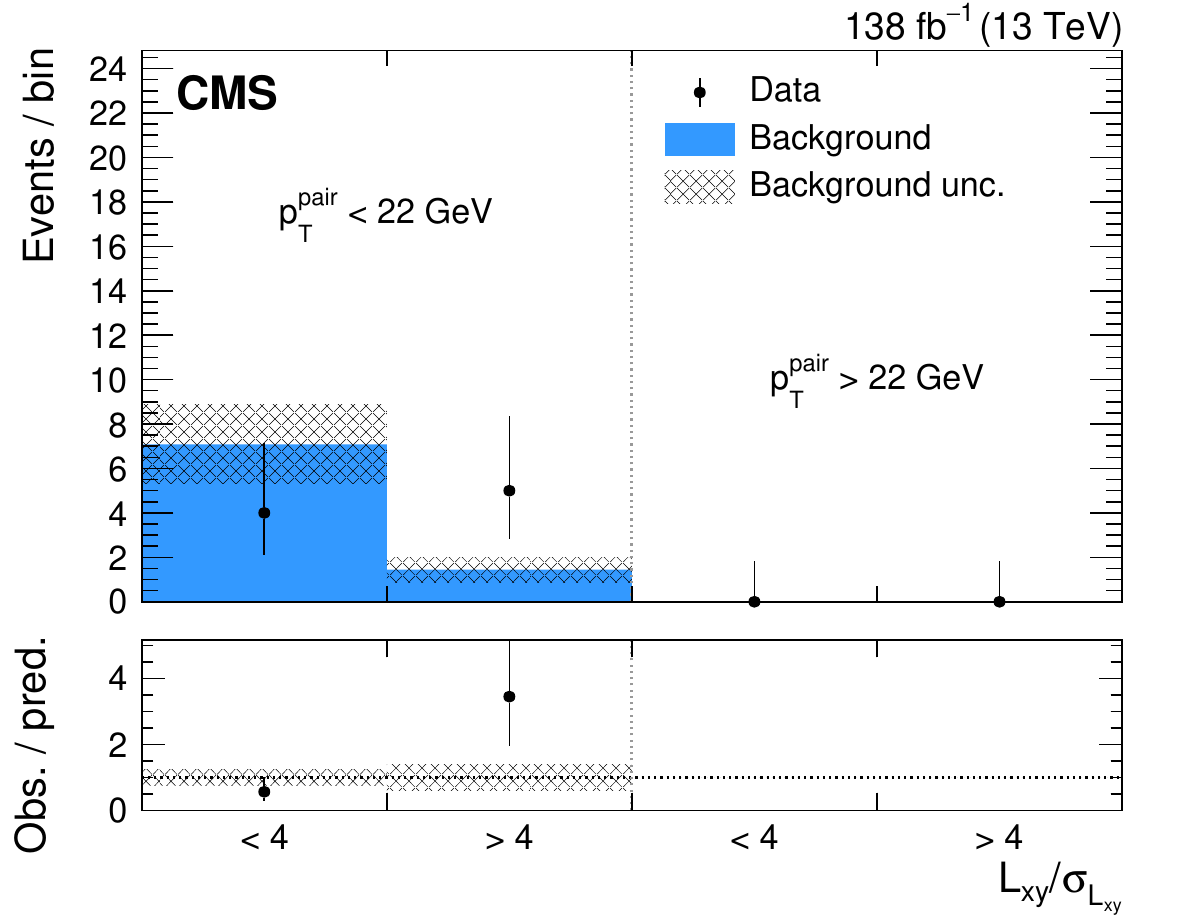}%
\hspace*{0.05\textwidth}%
\includegraphics[width=0.47\textwidth]{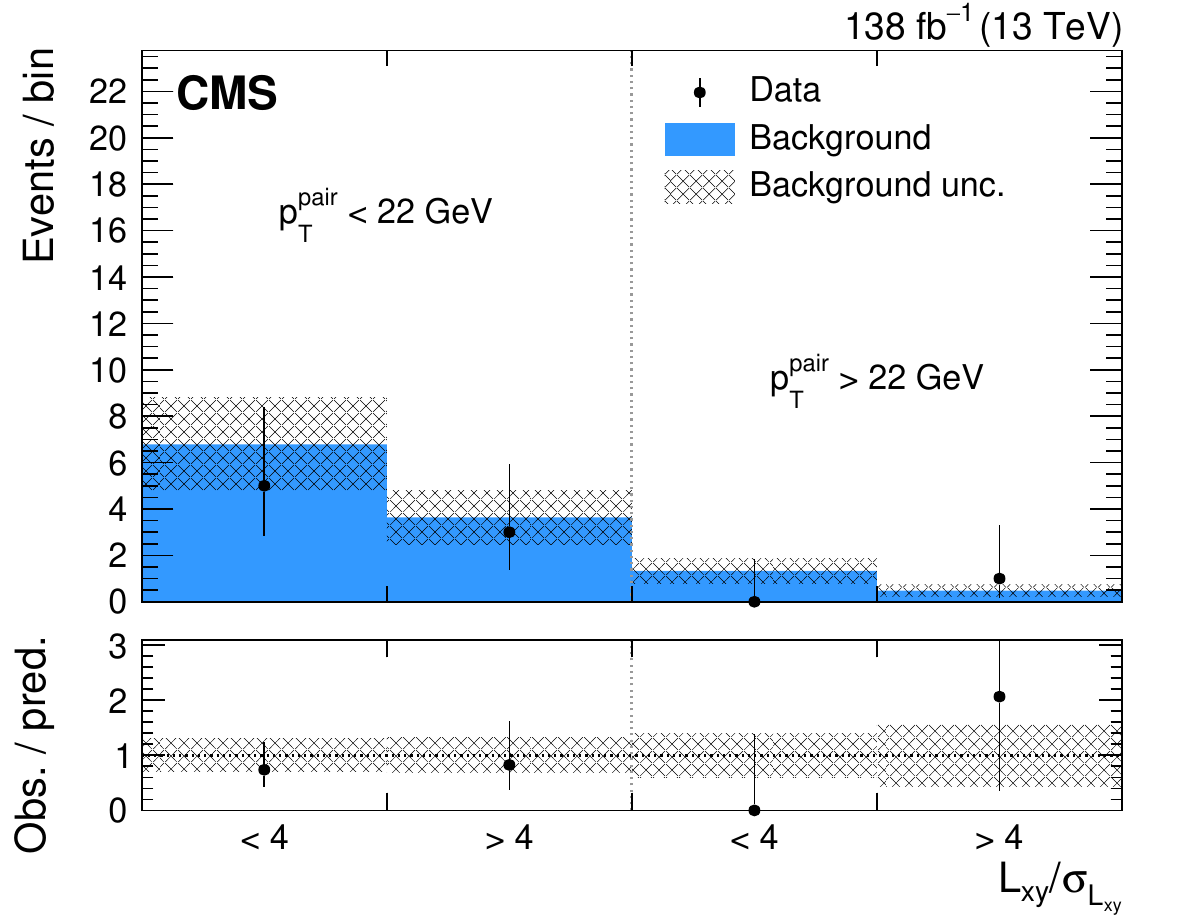}
\caption{%
    Observed (black points) and predicted (blue histograms) yields in the sideband with at least one non-isolated DSA muon of the low-\pt validation region (left) and the SR (right) in 2016--2018.
    The hatched bands represent the statistical uncertainty in the background prediction.
    The lower panels show the ratio between the observed and predicted yields in each bin.
    In the left figure, both predicted and observed yields are exactly zero in the last two bins by construction of the low-\pt validation region.
}
\label{fig:closure_invIso}
\end{figure}

To confirm the hypothesis that this background is negligible in the SR, we consider the sideband of the low-\pt validation region and of the SR where one or both of the DSA muons fails the \Ich requirement.
The results are shown in Fig.~\ref{fig:closure_invIso}.
In the sideband of the low-\pt validation region, an excess over the background prediction is observed for \mbox{bin 2}, similar to Fig.~\ref{fig:closure_lowPt} (left).
In contrast, good agreement is found in all bins in the sideband of the SR, indicating that this discrepancy is negligible in the high-\pt selection, consistent with the expectation that the excess is caused by photons converting into muons.
Nevertheless, we assign a systematic uncertainty for the observed non-closure in the low-\pt validation region, also considering the low statistical power of the validation with the inverted isolation requirement, as detailed in Section~\ref{sec:systematics}.

\subsection{Validation with inverted tracker veto}

\begin{table}[!t]
\centering
\topcaption{%
    Definition of the regions used in the ABCD method for the inverted tracker veto validation region.
}
\label{tab:abcd_CR_definition}
\renewcommand{\arraystretch}{1.2}
\begin{tabular}{lcc}
    \hline
    Region & Reduced signal selection & Veto muon $\chi^2$ \\
    \hline
    \Bpr & Pass & $>$1.5 \\
    \Bprpr & Pass & $<$1.5 \\
    \Dpr & Fail & $>$1.5 \\
    \Dprpr & Fail & $<$1.5 \\
    \hline
\end{tabular}
\end{table}

As an additional validation, we consider the inverted tracker veto (\ie the criterion used to define regions B and D) to define a CR.
Referring to the global or tracker muon that overlaps with \muone as \emph{veto muon}, we define a criterion on the \chisq of the track fit of the veto muon and use this together with the reduced signal selection to define a set of ABCD regions (labelled \Bpr, \Bprpr, \Dpr, \Dprpr) as shown in Table~\ref{tab:abcd_CR_definition}.
Events with $\chisq>1.5$ (regions \Bpr, \Dpr) have a poorer track quality compared to events with $\chisq<1.5$ (regions \Bprpr, \Dprpr), and thus mimic the background in the SR where no match with a veto muon is found.

\begin{figure}[!ht]
\centering
\includegraphics[width=0.47\textwidth]{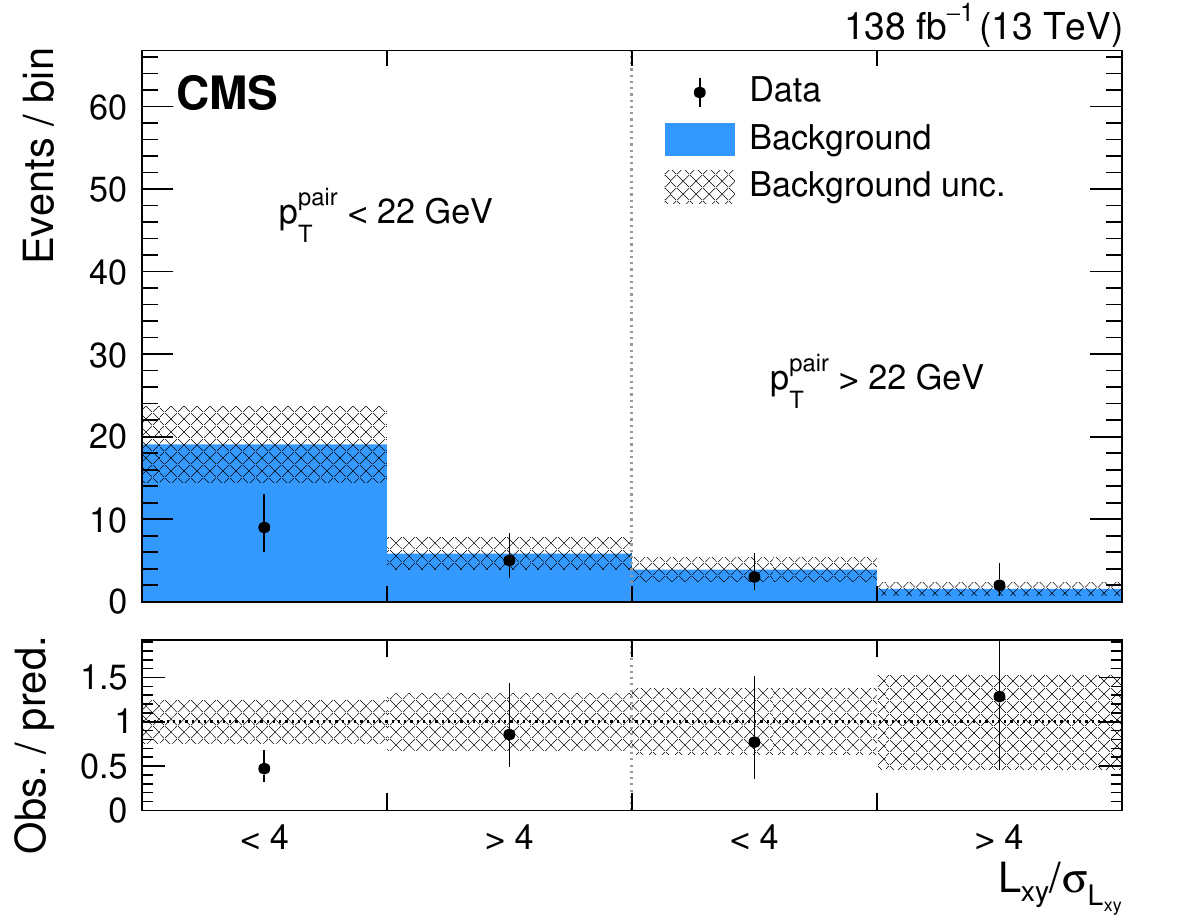}%
\hspace*{0.05\textwidth}
\includegraphics[width=0.47\textwidth]{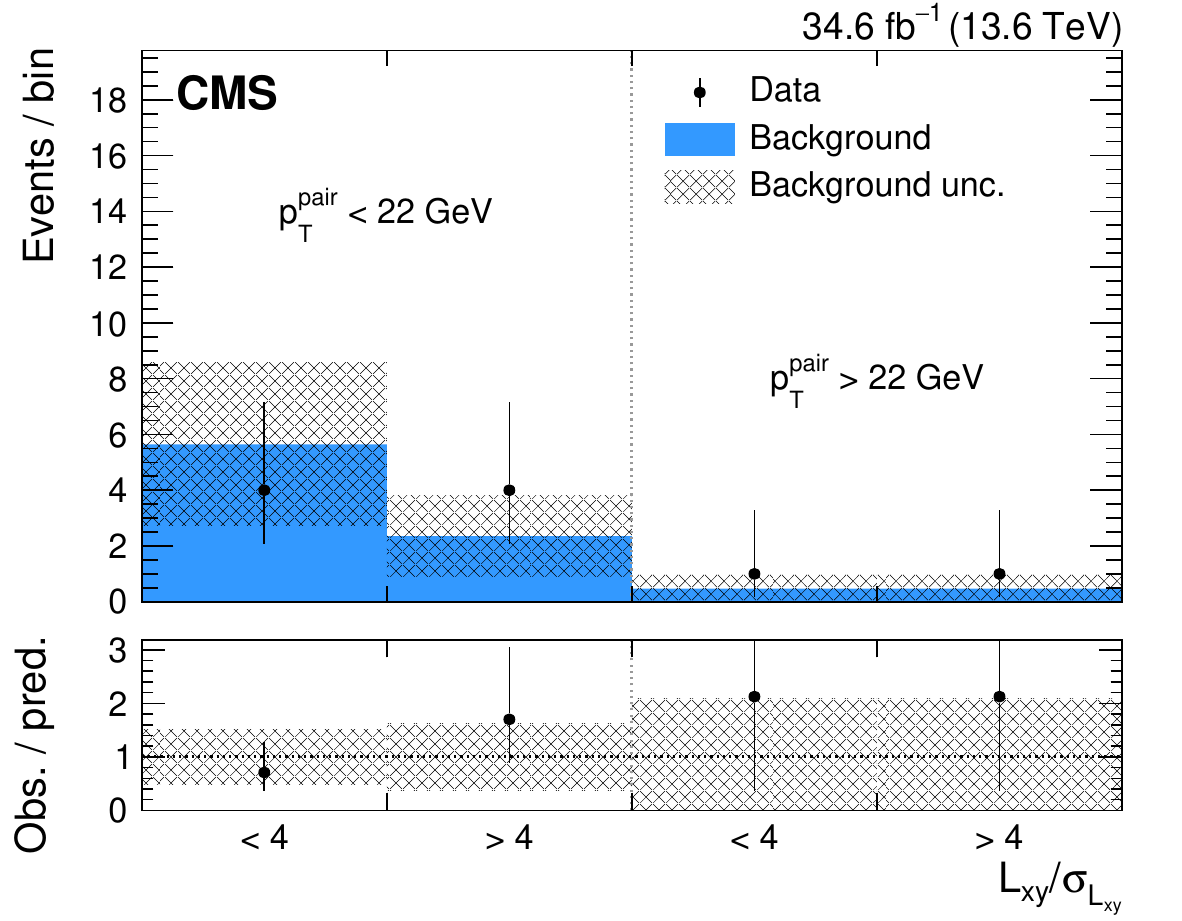}
\caption{%
    Observed (black points) and predicted (blue histograms) yields in the inverted tracker veto validation region in 2016--2018 (left) and in 2022 with relaxed DSA isolation requirements (right).
    The hatched bands represent the statistical uncertainty in the background prediction.
    The lower panels show the ratio between the observed and predicted yields in each bin.
}
\label{fig:closure_invVeto}
\end{figure}

The predicted yield of region \Bpr can then be estimated via the ABCD method as:
\begin{equation}
    \numBpr = \numBprpr \frac{\numDpr}{\numDprpr}.
\end{equation}
The results of the validation for the combined bins and with relaxed isolation criteria on the DSA muons are reported in Table~\ref{tab:abcd_yields_invVeto}, where the observed yield in region \Bpr is found in agreement with the prediction for all data-taking years.
Figure~\ref{fig:closure_invVeto} shows the comparison between observed and predicted yields in the four search bins, with relaxed DSA isolation requirements applied for 2022 data as in Fig.~\ref{fig:closure_lowPt}.

\begin{table}[!ht]
\centering
\topcaption{%
    Observed yields in the regions \Bpr, \Bprpr, \Dpr, and \Dprpr for the inverted tracker veto validation region with relaxed DSA isolation requirements, and the yield in region \Bpr predicted using the ABCD method.
}
\label{tab:abcd_yields_invVeto}
\renewcommand{\arraystretch}{1.2}
\begin{tabular}{lccccc}
    \hline
    Inverted tracker veto & \multicolumn{4}{c}{Observed} & Predicted \\[-3pt]
    validation region & \numBpr & \numBprpr & \numDpr & \numDprpr & \numBpr \\
    \hline
    2016--2018 & 49 & 149 & 37 & 95 & $58 \pm 12$  \\
    2022  & 10 & 19  & 8  & 17 & $8.9 \pm 4.3$  \\
    \hline
\end{tabular}
\end{table}

\section{Systematic uncertainties}
\label{sec:systematics}

\subsection{Systematic uncertainties in the signal yields}

Several sources of systematic uncertainty affect the expected signal yields estimated from the simulated samples.
Each uncertainty is evaluated separately for the four SR bins, except for the ones that only affect the overall signal normalization.
Each systematic uncertainty is then treated as fully correlated among the four SR bins in the fit.
Unless specifically stated, the systematic uncertainties in the 2016--2018 and 2022 signal yields are evaluated separately and are considered uncorrelated for the two data-taking periods.

While the MC generation of HNL signal events is performed at LO precision, we apply a scale factor to the signal normalization based on the ratio of NNLO and LO cross sections calculated for SM \PW boson production, similar to Ref.~\cite{CMS-EXO-20-009}.
Since higher-order QCD corrections primarily affect the \PW boson production from incoming partons rather than its decay, the NNLO/LO ratio is derived for $\PW\to\Pell\PGn$ events and applied to the HNL signal samples.
We then assign a systematic uncertainty in the signal normalization of 3.9\%, corresponding to the uncertainty in the NNLO cross section calculation, treated as correlated between 2016--2018 and 2022.
The statistical uncertainty in the signal yields from the limited number of simulated events in the MC samples is added as a systematic uncertainty, amounting to per-bin variations of 2.7--4\% in 2016--2018 and 3--6\% in 2022.

The integrated luminosity is measured with a precision of 1.2--2.5\% for the 2016--2018 data-taking years~\cite{CMS-lumi-2016, CMS-lumi-2017, CMS-lumi-2018}.
Because of the correlation between systematic effects across the different years, the integrated luminosity for the combined 2016--2018 data set is determined with an uncertainty of 1.6\%.
This value is included as a systematic source of uncertainty in the signal normalization for the 2016--2018 analysis.
The precision of the 2022 integrated luminosity measurement is 1.4\%~\cite{CMS-lumi-2022} and is accounted for as a systematic uncertainty in the signal normalization for the 2022 analysis.

The efficiency of the reconstruction and identification criteria used to select the final-state particles may be inaccurately modelled in the MC simulation used to estimate the signal yields, leading to discrepancies between the distributions obtained from MC samples and collision data.
To address this, efficiencies are separately computed in data and simulation, and data-to-MC scale factors are derived to correct the simulated samples for this effect.
The uncertainties in these scale factors introduce various sources of systematic uncertainty in the MC modelling of the signal.

Varying the signal yields by the uncertainties in the prompt-muon trigger, reconstruction, identification, and isolation scale factors results in changes of the signal yield by less than 1\% in all bins and for all cases.
Similarly, the uncertainties in the \PQb tagging efficiency scale factors, relevant for the \PQb jet veto, change the signal yields in all bins by less than 1\%.

In 2016--2018, given the limited time resolution of the CMS muon detectors, L1 trigger muons could be associated with the wrong bunch crossing, causing the loss of the event of interest---an effect known as L1 trigger prefiring.
Since this effect is not accounted for in the simulation, prefiring probabilities are separately computed for different physics objects, and event weights to correct the simulation are derived as the product of ``non-prefiring'' probabilities for all objects in the event.
The uncertainties in the prefiring weights are propagated to the signal yields, resulting in variations of less than 1\% in all bins.

The DSA muon efficiencies in the signal simulation are corrected using DSA reconstruction and identification scale factors first computed for prompt muons and then extrapolated to large displacements through studies of cosmic ray muon events.
We consider two sources of systematic uncertainties: the uncertainties in the DSA scale factors derived with prompt muons (computed as described above) and the uncertainty in the fit of the impact parameter of the DSA muons, $d_0$, which is used for the extrapolation.
The latter one is the dominant effect.
We treat the uncertainties in the scale factors applied to \muone and \mutwo as correlated between the two muons.
The overall systematic uncertainty in the signal yields from all scale factors applied for the displaced muons amounts to approximately 5.7\% in all bins.

Typical values of the systematic uncertainties in the four SR bins, from the sources described above, are shown in Table~\ref{table:syst_summary}.

\begin{table}[!ht]
\centering
\caption{Magnitude of the systematic uncertainties in the signal yields in the four SR bins. Values are given for an HNL signal corresponding to $\mN=3\GeV$ and $\VNmsq=4.9\times10^{-6}$, but no significant variations are observed for different signal hypotheses.}
\label{table:syst_summary}
\renewcommand{\arraystretch}{1.2}
\begin{tabular}{lcccc}
    \hline
    \multirow{2}{*}{Systematic source} & \multicolumn{4}{c}{Relative systematic uncertainty {[}\%{]}} \\[-2pt]
    & \makebox[1.2cm]{Bin 1} & \makebox[1.2cm]{Bin 2} & \makebox[1.2cm]{Bin 3} & \makebox[1.2cm]{Bin 4} \\
    \hline
    Signal normalization                & 3.9 & 3.9 & 3.9 & 3.9 \\
    MC event count                      & 4.0 & 3.5 & 3.2 & 2.7 \\
    Integrated luminosity               & 1.6 & 1.6 & 1.6 & 1.6 \\
    Prompt \PGm trigger                 & 0.1 & 0.1 & 0.1 & 0.1 \\
    Prompt \PGm reconstruction          & 0.1 & 0.1 & 0.1 & 0.1 \\
    Prompt \PGm identification          & 0.2 & 0.2 & 0.2 & 0.2 \\
    Prompt \PGm isolation               & 0.2 & 0.2 & 0.1 & 0.1 \\
    \PQb tagging scale factors          & 0.1 &\makebox[0pt][r]{$<$}0.1& 0.1 & 0.1 \\
    L1 prefiring weights                & 0.1 & 0.1 & 0.1 & 0.1 \\
    DSA \PGm identification: prompt     & 0.6 & 0.6 & 0.6 & 0.6 \\
    DSA \PGm reconstruction: high $d_0$ & 4.0 & 4.0 & 4.0 & 4.0 \\
    DSA \PGm identification: high $d_0$ & 4.0 & 4.0 & 4.0 & 4.0 \\
    \hline
\end{tabular}
\end{table}

\subsection{Systematic uncertainties in the background yields}

Three sources of systematic uncertainty are included for the background yields.
First, the uncertainty in the \TF, which is dominated by \numC, is added as a single source of uncertainty in the background normalization, and amounts to 26\% in 2016--2018 and 49\% in 2022.
The second source is the statistical uncertainty in \numB, which is propagated to the SR bins by using a Gamma distribution, with a shape parameter given by the \numB distribution and a scale parameter by \TF.

Finally, an additional uncertainty is assigned to account for the non-closure observed in the low-\pt validation region and for the limited statistical power of the validation at high \pt.
The four per-bin uncertainties used in the SR are derived from the 2016--2018 VR with the inverted DSA isolation requirement (Fig.~\ref{fig:closure_invIso}, right).
In each bin $i$, the observed yield \numA in this sideband is compared with the ABCD prediction $\numB\TF$. The central 68\% interval of the ratio $\numA/(\numB\TF)$ is determined using pseudoexperiments in which \numA and \numB are varied according to Poisson distributions, while the \TF is assumed to be known with high precision. The resulting interval is used to define the systematic uncertainty in each bin. The largest relative uncertainty is obtained in \mbox{bin 4}, owing to the small number of events in the validation region for this bin.
The relative values of this systematic uncertainty derived from 2016--2018 data are also applied to the 2022 background yields, where similar physics effects are expected but the limited number of events in the VR prevents a precise estimate of the corresponding uncertainty.
The absolute uncertainty in \mbox{bin 4} amounts to 3 and 0.9 events in 2016--2018 and 2022, respectively.

\section{Results and interpretation}
\label{sec:results}

\begin{figure}[!ht]
\centering
\includegraphics[width=0.47\textwidth]{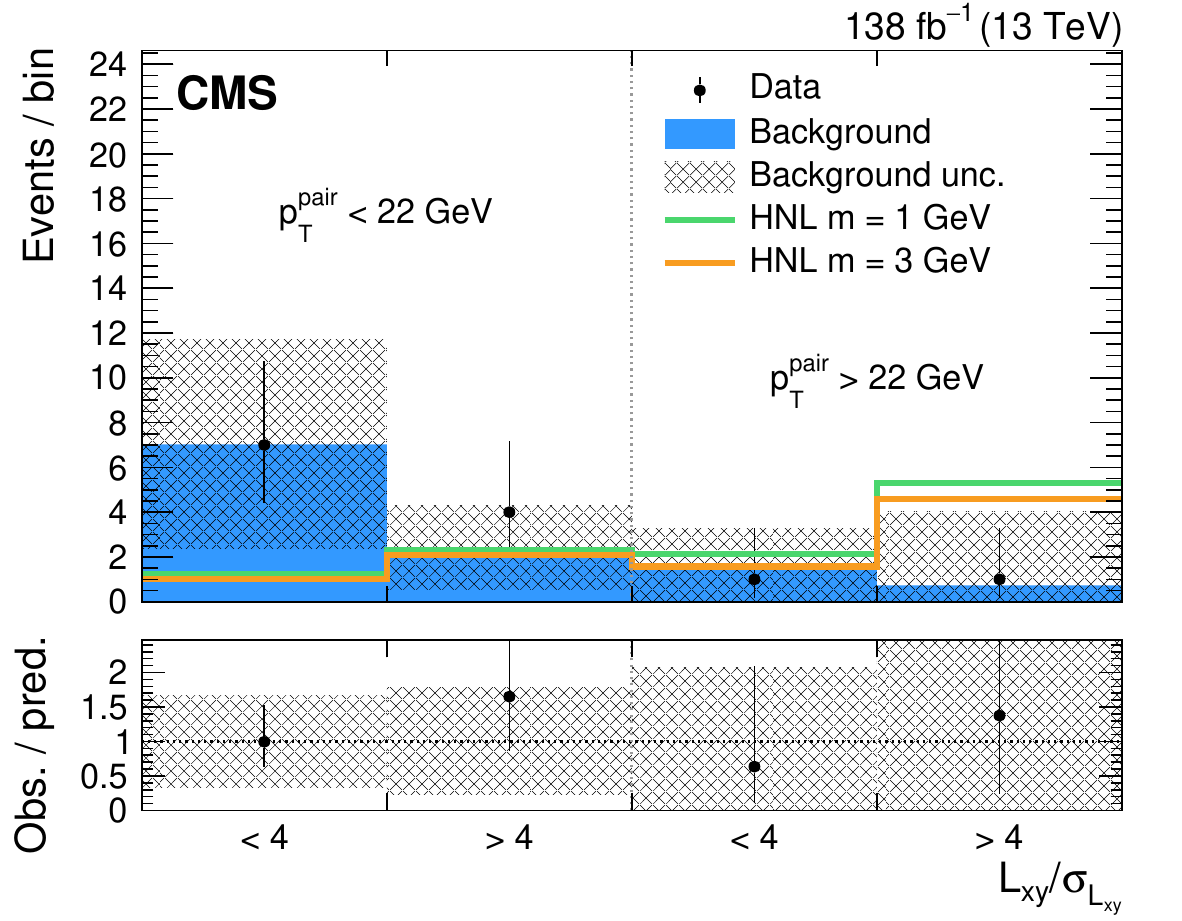}
\hspace*{0.05\textwidth}%
\includegraphics[width=0.47\textwidth]{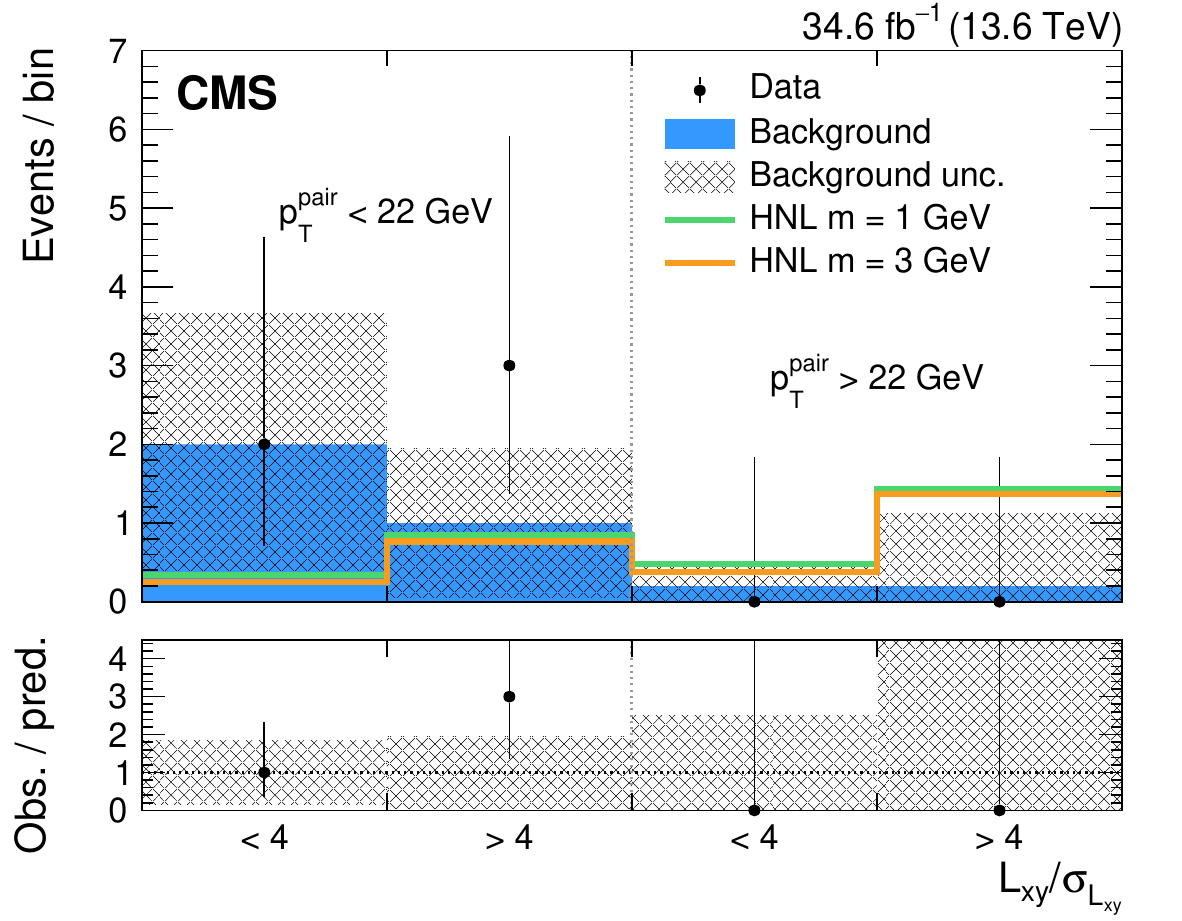}
\caption{%
    Observed (black points) and predicted (blue histograms) yields in the SR bins in 2016--2018 (left) and 2022 (right).
    The hatched bands represent the total uncertainty in the background prediction.
    The green and orange histograms show the signal distributions for two different hypotheses, corresponding to $\mN=1$ and 3\GeV, and respective \VNmsq values equal to $1.4\times10^{-4}$ and $4.9\times10^{-6}$, corresponding to the smallest excluded values at 95\% CL in Ref.~\cite{CMS-EXO-20-009}.
    The lower panels show the ratio between the observed yields and predicted background yields in each bin.
}
\label{fig:bkg_expected_SR}
\end{figure}

The observed yields in the four SR bins, together with the corresponding background predictions, are shown in Fig.~\ref{fig:bkg_expected_SR}, separately for the 2016--2018 and 2022 data sets.
No significant deviation with respect to the background-only hypothesis is found.

We assume the Majorana HNL hypothesis and derive exclusion limits at 95\% \CL on several signal scenarios using the modified frequentist \CLs approach~\cite{Junk:1999kv, CLs_technique}.
For a given signal hypothesis, a binned likelihood function $L(r,\theta)$ is constructed as the product of Poisson probabilities to obtain the observed data given the expected signal rates scaled by the parameter $r$, referred to as signal strength, and the background prediction in the four analysis bins.
The different sources of systematic uncertainty in the estimated signal and background yields described in Section~\ref{sec:systematics} are included in $L$ as nuisance parameters, collectively indicated by the symbol $\theta$.
The compatibility between the observed data and a given value of $r$ is estimated through the LHC test statistic $q(r)=-2\ln L(r,\hat{\theta}_r)/L(\hat{r},\hat{\theta})$ for $0\leq\hat{r}\leq r$, where $\hat{\theta}_r$ is the maximum likelihood estimator of $\theta$ for a given $r$, while $\hat{r}$ and $\hat{\theta}$ are the estimators corresponding to the global maximum of $L$~\cite{ATLAS:2011tau}.
We exclude the considered signal scenario if the obtained \CLs upper limit on the signal strength $r$ at 95\% \CL is lower or equal to 1, \ie if the value of $r=1$ is incompatible with the observed data at 95\% \CL.
The statistical evaluation has been performed with the CMS statistical analysis tool \textsc{combine}~\cite{CAT-23-001}, which is based on the \textsc{RooFit}~\cite{Verkerke:2003ir} and \textsc{RooStats}~\cite{Moneta:2010pm} frameworks.

\begin{figure}[!b]
\centering
\includegraphics[width=0.5\textwidth]{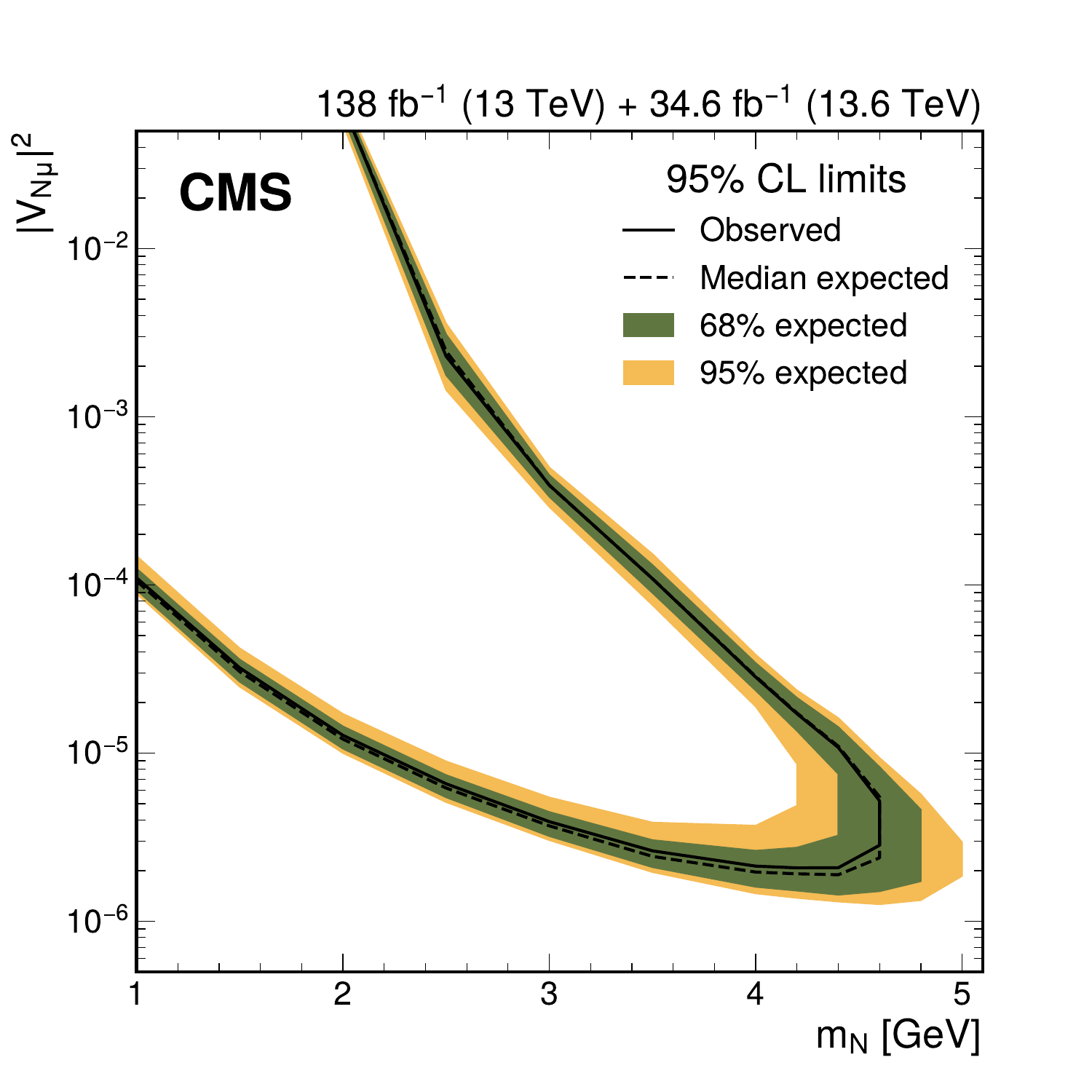}
\caption{%
    Exclusion limits at 95\% \CL on \VNmsq as functions of \mN for a Majorana HNL, obtained with the combined data set.
    The solid (dashed) black curve indicates the observed (expected) exclusion, where the parameter combinations inside the curve are excluded.
    The green and yellow bands indicate the 68 and 95\% confidence intervals around the median expected exclusion, respectively.
}
\label{fig:limits_m_v2_Majorana}
\end{figure}

\begin{figure}[!ht]
\centering
\includegraphics[width=0.5\textwidth]{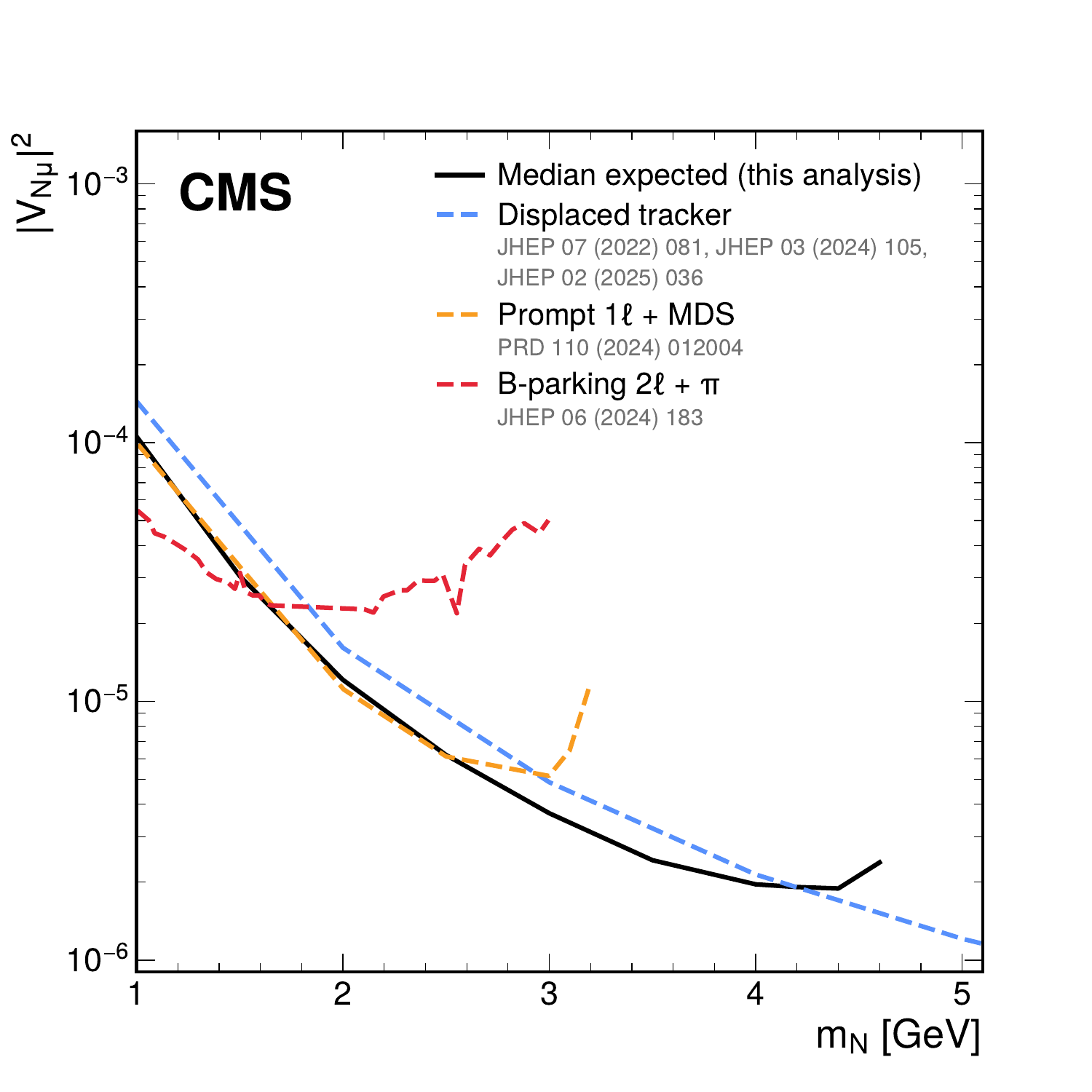}
\caption{%
    Lower boundary of the median expected excluded region at 95\% \CL in \VNmsq and \mN for a Majorana HNL, obtained with the combination of 2016--2018 and 2022 data (solid black line), compared with the expected limits obtained by other CMS analyses in the same mass range (dashed coloured lines)~\cite{CMS-EXO-20-009,CMS:2023jqi,CMS:2024hik,CMS:2024ake,CMS:2024ita}.
}
\label{fig:limits_m_v2_Majorana_comparison}
\end{figure}

Exclusion limits are derived on \VNmsq as functions of \mN in the scenario of one Majorana HNL, using a grid of points in the $(\mN,\VNmsq)$ parameter space.
Given the effect of \VNmsq on the analysis acceptance and the final kinematic distributions used for the binning, a simple rescaling of the upper limits on $r$ is not sufficient to find the excluded values of \VNmsq for a given \mN value.
Hence, signal rates in each SR bin are derived for different values of \VNmsq using the lifetime-based weights described in Section~\ref{sec:signal_simulation} starting from the available MC samples at a given mass, and upper limits on $r$ are separately computed for each $(\mN,\VNmsq)$ point.
The excluded region in the $(\mN,\VNmsq)$ parameter space is shown in Fig.~\ref{fig:limits_m_v2_Majorana}.
The loss of sensitivity above $\mN\approx4.5\GeV$ is driven
primarily by the HNL lifetime dependence on \mN. At fixed values of \VNmsq, the lifetime decreases as $\mN^{-5}$, so heavier HNLs increasingly decay before reaching the region targeted by this search. Smaller mixing values not only increase the decay length, but also suppress the production rate, leaving no parameter region with sufficient
signal yield at higher masses.
To compare the sensitivity at long HNL lifetimes with that of previous searches, Fig.~\ref{fig:limits_m_v2_Majorana_comparison} shows the lower boundary of the median expected excluded region obtained in this analysis and by other CMS analyses targeting HNL decays in a similar mass range. The upper boundary is not included in this comparison because the corresponding region at large \VNmsq is already excluded by previous searches.
The expected limits for low \VNmsq values obtained in this analysis from the combination of the data sets exceed the expected limits of previous analyses for masses
between 2.5 and 4.2\GeV.

\section{Summary}
\label{sec:summary}

A search for heavy neutral leptons (HNLs) has been performed using proton-proton collision events with one prompt muon and two highly displaced muons.
The analyzed data set was recorded by the CMS experiment at the LHC and corresponds to an integrated luminosity of 138\fbinv recorded at $\sqrt{s}=13\TeV$ in 2016--2018 and 34.6\fbinv at 13.6\TeV in 2022.
The target process is HNL production in association with a prompt muon from the decay of a \PW boson, with a delayed HNL decay to two muons and one standard model neutrino.
The two displaced muons are reconstructed using only information from the CMS muon system, thus significantly extending the sensitivity of the analysis to HNLs with longer decay lengths compared to previous searches for long-lived HNLs.
No significant deviation from the standard model background estimated from control samples in data is found.
Under the Majorana HNL hypothesis, exclusion limits at 95\% confidence level are set on the HNL mixing parameter as functions of its mass.
This analysis surpasses previous experimental results in terms of the sensitivity to HNLs with masses between 2.5 and 4.2\GeV.

\begin{acknowledgments}
We congratulate our colleagues in the CERN accelerator departments for the excellent performance of the LHC and thank the technical and administrative staffs at CERN and at other CMS institutes for their contributions to the success of the CMS effort. In addition, we gratefully acknowledge the computing centres and personnel of the Worldwide LHC Computing Grid and other centres for delivering so effectively the computing infrastructure essential to our analyses. Finally, we acknowledge the enduring support for the construction and operation of the LHC, the CMS detector, and the supporting computing infrastructure provided by the following funding agencies: SC (Armenia), BMFWF and FWF (Austria); FNRS and FWO (Belgium); CNPq, CAPES, FAPERJ, FAPERGS, and FAPESP (Brazil); MES and BNSF (Bulgaria); CERN; CAS, MoST, and NSFC (China); MINCIENCIAS (Colombia); MSES and CSF (Croatia); RIF (Cyprus); SENESCYT (Ecuador); ERC PRG and PSG, TARISTU24-TK10 and MoER TK202 (Estonia); Academy of Finland, MEC, and HIP (Finland); CEA and CNRS/IN2P3 (France); SRNSF (Georgia); BMFTR, DFG, and HGF (Germany); GSRI (Greece); MATE and NKFIH (Hungary); DAE and DST (India); IPM (Iran); SFI (Ireland); INFN (Italy); MSIT and NRF (Republic of Korea); MES (Latvia); LMTLT (Lithuania); MOE and UM (Malaysia); BUAP, CINVESTAV, CONACYT, LNS, SEP, and UASLP-FAI (Mexico); MOS (Montenegro); MBIE (New Zealand); PAEC (Pakistan); MSHE, NSC, and NAWA (Poland); FCT (Portugal); MESTD (Serbia); MICIU/AEI and PCTI (Spain); MOSTR (Sri Lanka); Swiss Funding Agencies (Switzerland); NSTC (Taiwan); MHESI (Thailand); TUBITAK and TENMAK (T\"{u}rkiye); NASU (Ukraine); STFC (United Kingdom); DOE and NSF (USA).

\begin{sloppypar}
\setlength\emergencystretch{\hsize}
\hyphenation{Rachada-pisek} Individuals have received support from the Marie-Curie programme and the European Research Council and Horizon 2020 Grant, contract Nos.\ 675440, 724704, 752730, 758316, 765710, 824093, 101115353, 101002207, 101001205, and COST Action CA16108 (European Union); the Leventis Foundation; the Alfred P.\ Sloan Foundation; the Alexander von Humboldt Foundation; the Science Committee, project no. 22rl-037 (Armenia); the Fonds pour la Formation \`a la Recherche dans l'Industrie et dans l'Agriculture (FRIA) and Fonds voor Wetenschappelijk Onderzoek contract No. 1228724N (Belgium); the Beijing Municipal Science \& Technology Commission, No. Z191100007219010, the Fundamental Research Funds for the Central Universities, the Ministry of Science and Technology of China under Grant No. 2023YFA1605804, the Natural Science Foundation of China under Grant No. 12535004, and USTC Research Funds of the Double First-Class Initiative No.\ YD2030002017 (China); the Ministry of Education, Youth and Sports (MEYS) of the Czech Republic; the Shota Rustaveli National Science Foundation (Georgia); the Deutsche Forschungsgemeinschaft (DFG), among others, under Germany's Excellence Strategy -- EXC 2121 ``Quantum Universe" -- 390833306, and under project number 400140256 - GRK2497; the Hellenic Foundation for Research and Innovation (HFRI), Project Number 2288 (Greece); the Hungarian Academy of Sciences, the New National Excellence Program - \'UNKP, the NKFIH research grants K 131991, K 138136, K 143460, K 143477, K 147557, K 146913, K 146914, K 147048, TKP2021-NKTA-64, 2026-4.1.1-MISSZI\'{O}K-2026-00005 and 2025-1.1.5-NEMZ\_KI-2025-00004, and MATE KKP and KKPCs Research Excellence and Flagship Research Groups grants (Hungary); the Council of Science and Industrial Research, India; ICSC -- National Research Centre for High Performance Computing, Big Data and Quantum Computing, FAIR -- Future Artificial Intelligence Research, and CUP I53D23001070006 (Mission 4 Component 1), funded by the NextGenerationEU program, the Italian Ministry of University and Research (MUR) under Bando PRIN 2022 -- CUP I53C24002390006, PRIN PRIMULA 2022RBYK7T (Italy); the Latvian Council of Science; the Ministry of Science and Higher Education, project no. 2022/WK/14, and the National Science Centre, contracts Opus 2021/41/B/ST2/01369, 2021/43/B/ST2/01552, 2023/49/B/ST2/03273, and the NAWA contract BPN/PPO/2021/1/00011 (Poland); the Funda\c{c}\~ao para a Ci\^encia e a Tecnologia (Portugal); the National Priorities Research Program by Qatar National Research Fund; MICIU/AEI/10.13039/501100011033, ERDF/EU, ``European Union NextGenerationEU/PRTR", projects PID2022-142604OB-C21, PID2022-139519OB-C21, PID2023-147706NB-I00, PID2023-148896NB-I00, PID2023-146983NB-I00, PID2023-147115NB-I00, PID2023-148418NB-C41, PID2023-148418NB-C42, PID2023-148418NB-C43, PID2023-148418NB-C44, PID2024-158190NB-C22, RYC2021-033305-I, RYC2024-048719-I, CNS2023-144781, CNS2024-154769 and Plan de Ciencia, Tecnolog{\'i}a e Innovaci{\'o}n de Asturias, Spain; the Chulalongkorn Academic into Its 2nd Century Project Advancement Project, the National Science, Research and Innovation Fund program IND\_FF\_68\_369\_2300\_097, and the Program Management Unit for Human Resources \& Institutional Development, Research and Innovation, grant B39G680009 (Thailand); the Eric \& Wendy Schmidt Fund for Strategic Innovation through the CERN Next Generation Triggers project under grant agreement number SIF-2023-004; the Kavli Foundation; the Nvidia Corporation; the SuperMicro Corporation; the Welch Foundation, contract C-1845; and the Weston Havens Foundation (USA).
\end{sloppypar}
\end{acknowledgments}\section*{Data availability} Release and preservation of data used by the CMS Collaboration as the basis for publications is guided by the  \href{https://doi.org/10.7483/OPENDATA.CMS.1BNU.8V1W}{CMS data preservation, re-use and open access policy}.

\bibliography{auto_generated}
\cleardoublepage \appendix\section{The CMS Collaboration \label{app:collab}}\begin{sloppypar}\hyphenpenalty=5000\widowpenalty=500\clubpenalty=5000\cmsinstitute{Yerevan Physics Institute, Yerevan, Armenia}
{\tolerance=6000
A.~Gevorgyan\cmsorcid{0000-0003-2751-9489}, A.~Hayrapetyan, V.~Makarenko\cmsorcid{0000-0002-8406-8605}, A.~Tumasyan\cmsAuthorMark{1}\cmsorcid{0009-0000-0684-6742}
\par}
\cmsinstitute{Institut f\"{u}r Hochenergiephysik, Vienna, Austria}
{\tolerance=6000
P.S.~Hussain\cmsorcid{0000-0002-4825-5278}, M.~Sonawane\cmsorcid{0000-0003-0510-7010}
\par}
\cmsinstitute{Marietta Blau Institute for Particle Physics, Vienna, Austria}
{\tolerance=6000
W.~Adam\cmsorcid{0000-0001-9099-4341}, L.~Benato\cmsorcid{0000-0001-5135-7489}, T.~Bergauer\cmsorcid{0000-0002-5786-0293}, M.~Dragicevic\cmsorcid{0000-0003-1967-6783}, S.~Gundacker\cmsorcid{0000-0003-2087-3266}, A.K.~Guven\cmsorcid{0009-0004-5670-5138}, M.~Jeitler\cmsAuthorMark{2}\cmsorcid{0000-0002-5141-9560}, N.~Krammer\cmsorcid{0000-0002-0548-0985}, A.~Li\cmsorcid{0000-0002-4547-116X}, D.~Liko\cmsorcid{0000-0002-3380-473X}, M.~Matthewman, A.~Pfeiffer\cmsorcid{0000-0001-5328-448X}, J.~Schieck\cmsAuthorMark{2}\cmsorcid{0000-0002-1058-8093}, R.~Sch\"{o}fbeck\cmsAuthorMark{2}\cmsorcid{0000-0002-2332-8784}, M.~Shooshtari\cmsorcid{0009-0004-8882-4887}, N.~Van~Den~Bossche\cmsorcid{0000-0003-2973-4991}, W.~Waltenberger\cmsorcid{0000-0002-6215-7228}, C.E.~Wulz\cmsAuthorMark{2}\cmsorcid{0000-0001-9226-5812}
% ...
\par}
$^{1}$Also at Yerevan State University, Yerevan, Armenia\\
$^{2}$Also at Technische Universit\"{a}t Wien, Vienna, Austria\\
% ...
\end{sloppypar}
\end{document}